\documentclass[11pt]{article}
\usepackage{epsfig,a4}
\def\be{\begin{equation}}
\def\ee{\end{equation}}
\def\bea{\begin{eqnarray}}
\def\eea{\end{eqnarray}}

\def\C{{\rm\kern.24em
    \vrule width.02em height1.4ex depth-.05ex
    \kern-.26em C}}
\def\N{{\rm I\kern-.18em N}}
\def\R{{\rm I\kern-.21em R}}
\def\Z{{\rm\kern.26em
    \vrule width.02em height0.5ex depth 0ex
    \kern.04em
    \vrule width.02em height1.47ex depth-1ex
    \kern-.34em Z}}
\def\d{{\rm\kern.22em
    \vrule width.02em height1.0ex depth0ex
    \kern-.24em d}}

\def\fr{\frac}

\newcommand\qs{\!\not \! q}
\def\del{\partial}
\def\gam{\gamma}

\newcommand\qt{\tilde{q}}

\newcommand\real{\Re\mbox{e}\,}
\newcommand\imag{\Im\mbox{m}\,}

\newcommand\lbr{\left(}
\newcommand\rbr{\right)}
\newcommand\lbk{\left[}
\newcommand\rbk{\right]}

\def\l.{\left.}
\def\r.{\right.}
\newcommand\bmat{\boldmath}
\newcommand\ubmat{\unboldmath}

\newdimen\picraise
\newcommand\picbox[1]
{
  \setbox0=\hbox{\input{#1}}
  \picraise=-0.5\ht0 
  \advance\picraise by 0.5\dp0
  \advance\picraise by 3pt     
  \hbox{\raise\picraise \box0}
}
\newdimen\picraiset
\newcommand\picding[1]
{
  \setbox0=\hbox{\input{#1}}
  \picraiset=-0.5\ht0 
  \advance\picraiset by 0.5\dp0
  \advance\picraiset by -3pt  
  \hbox{\raise\picraiset \box0}
}
\newdimen\picraisehallo
\newcommand\pichallo[2]
{
  \setbox0=\hbox{\input{#1}}
  \picraisehallo=-0.5\ht0 
  \advance\picraisehallo by 0.5\dp0
  \advance\picraisehallo by -#2pt  
  \hbox{\raise\picraisehallo \box0}
}
\begin{document}
\begin{titlepage}
\begin{flushright}
Cavendish-HEP-99/11\\
DAMTP-1999-148\\
hep-ph/0001038
\\
\end{flushright}
\vfill
\begin{center}
\boldmath
{\LARGE{\bf Gribov's Equation for the Green Function}}\\[.2cm]
{\LARGE{\bf of Light Quarks$^*$}}
\unboldmath
\end{center}
\vspace{1.5cm}
\begin{center}
{\bf \Large
Carlo Ewerz
}
\end{center}
\vspace{.2cm}
\begin{center}
{\sl
Cavendish Laboratory, Cambridge University\\
Madingley Road, Cambridge CB3 0HE, UK\\[.2cm]}
and\\[.2cm]
{\sl 
DAMTP, Centre for Mathematical Sciences, Cambridge University\\ 
Wilberforce Road, Cambridge CB3 0WA, UK\\[.4cm]}
email: {\sl carlo@hep.phy.cam.ac.uk}\\[.4cm]
\end{center}
\vfill
\begin{abstract}
Gribov's scenario of supercritical charges in QCD is investigated. 
We perform a numerical study of the corresponding 
equation for the Green function of light quarks. This is done 
in an approximation which neglects all pion contributions. 
Different types of solutions in the Euclidean region 
are discussed and the mass function of the quark is calculated. 
The solutions of the equation are shown to have a qualitatively 
different behaviour if the strong coupling constant $\alpha_s$ 
exceeds a critical value $\alpha_c = 0.43$ in the infrared region. 
Chiral symmetry breaking is found to occur 
at supercritical coupling. The analytic structure of the solutions 
is investigated. Earlier results obtained by Gribov are confirmed and 
extended. 
\vfill
\end{abstract}
\vspace{5em}
\hrule width 5.cm
\vspace*{.5em}
{\small \noindent 
$^*$Work supported in part by the EU Fourth Framework Programme
`Training and Mobility of Researchers', Network `Quantum Chromodynamics
and the Deep Structure of Elementary Particles',
contract FMRX-CT98-0194 (DG 12 - MIHT).
}
\end{titlepage}

\section{Introduction}
\label{intro}
The breaking of chiral symmetry and the confinement of 
quarks and gluons are two of the most important 
properties of QCD. The details of the mechanism leading 
to confinement are still largely unknown, and the understanding 
of non--perturbative dynamics in QCD in general is still rather poor. 
A new picture of the confinement mechanism 
and of chiral symmetry breaking was developed by 
V.\,N.\ Gribov \cite{physica}--\cite{EPJ2}. It is based on the 
phenomenon of supercritical charges which can occur in QCD due to 
the existence of very light quarks. Its consequence is a dramatic 
change in the vacuum structure of the light quarks compared 
to the usual perturbative picture at small coupling. 

The phenomenon of supercritical charges is well--known in QED 
(for an extensive  review see \cite{Greiner}). The energy of 
the bound--state levels in the field of an isolated heavy nucleus 
decreases if the charge $Z$ of the nucleus is increased. 
When the charge exceeds 
a critical value\footnote{This number holds for a point--like nucleus, 
for an extended charge it is around $Z_{cr} \sim 165 $.} 
of $Z_{cr}=137$, the lowest bound--state level dives into 
the Dirac sea, i.\,e.\ sinks below $-m_{\mathrm e}$. As a consequence 
an electron from the (filled) continuum undergoes a transition into 
this level, and a positron is emitted. 
The electron is said to `fall onto the center'. 
In this situation the simple quantum mechanical picture breaks down, 
and the emerging bound state is in fact a collective state with a high 
probability to find an electron very close to the nucleus. 
This mechanism is called supercritical binding. 
The condition for its occurrence is that the Compton wavelength 
$1/m$ of the electron is much larger than the radius of the heavy 
charge. 

Gribov's confinement scenario is based on the idea that a similar 
phenomenon occurs in QCD due to the existence of very light 
(almost massless) quarks. The crucial point is that in this 
scenario already the color charge of a single quark is supercritical. 
Since this applies also to the light quarks themselves the  
situation is more involved than in QED. We will give only 
a condensed description of the resulting scenario here. More 
detailed accounts have been given in \cite{Lund,EPJ2,Orsay,GribovErice}. 
In order to get an understanding of the underlying physical 
picture we again use the quantum mechanical description, having 
in mind that the quantitative analysis should be 
based on the full underlying quantum field theory. 

The confinement of heavy quarks in Gribov's scenario is very 
similar to the supercritical binding in QED. Due to its supercritical 
charge the heavy quark captures a light antiquark from the vacuum, 
thus decaying into a supercritical heavy--light bound state. At the same 
time a light quark is created. This light quark decays again, as 
we will discuss now. 

\begin{figure}[htbp]
\begin{center}
\input{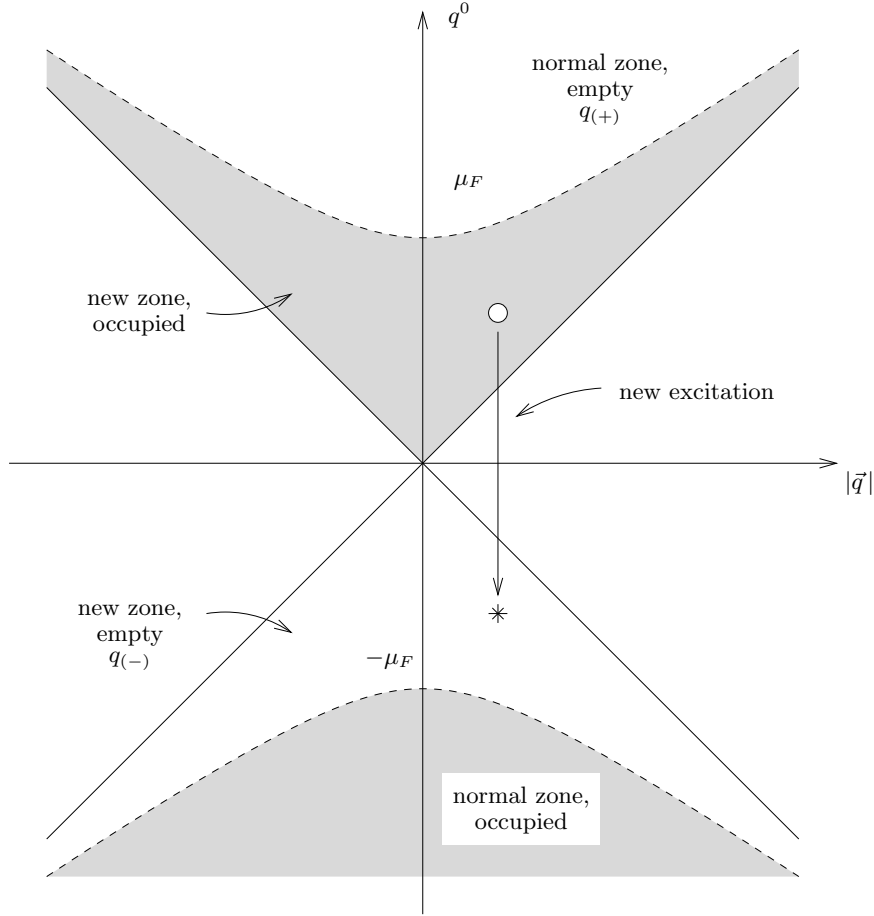}
\end{center}
\caption{Vacuum of light quarks in Gribov's scenario. 
  $q^0$ and $\vec{q}$ denote the (kinetic) energy and 
 three--momentum, respectively. }
\label{fig:newvac}
\end{figure}

In order to understand the confinement of light quarks, 
we first consider a bound state of a light quark and 
a light antiquark. If the coupling constant is small this is 
just a normal bound state like positronium in which the 
quarks have positive kinetic energy. If we now increase 
the coupling the binding energy will also increase. The total 
energy of this state will thus decrease. But if the coupling is 
further increased --- and becomes supercritical --- 
a situation is possible in which the total energy of the bound 
state becomes negative. In order to have a stable vacuum, however, 
the existence of negative energy particles has to be avoided. 
Consequently, the corresponding quark 
states in the supercritical bound states have to be filled in the vacuum.  
Therefore there are filled quark states with positive kinetic 
energy in addition to the usual filled states in the 
Dirac sea (see\footnote{In 
this figure only the energy of the quark is shown. One has 
to keep in mind that some of the states shown here exist only 
within supercritical bound states with an antiquark as 
described above.} Fig.\ \ref{fig:newvac}). 
The scale $\mu_F$ separating filled and empty states of 
positive kinetic energy resembles the Fermi surface 
in solid state physics. 
The existence of the additional states in the vacuum 
of light quarks implies also the 
existence of additional excitations of this vacuum 
which have quite unusual properties. The $q \bar{q}$ 
pair of such an excitation forms a 
supercritical bound state in which the quark 
and antiquark both have negative kinetic energy. 
They are interacting repulsively, and the supercritical bound 
state has positive total energy. The `binding force' 
leading to this unusual meson\footnote{Some possible 
properties of these novel mesons 
have been discussed in \cite{a980,YuriErice}.} is 
the Pauli exclusion principle. 
The quark and antiquark are bound in this meson because 
all other energetically possible states in the vacuum are 
already filled. 

Having discussed the emergence of the 
novel meson states, we can now understand the confinement 
of light quarks. According to Gribov it is caused by 
the continuous decay of the light quark. 
Any quark (with positive or negative kinetic energy, $q_{(+)}$ 
or $q_{(-)}$) decays into a supercritical bound state $M$ 
and a quark $q_{(-)}$ of negative kinetic energy, 
\be
q_{(\pm)} \rightarrow M + q_{(-)} \,.
\ee
In this sense the quark exists only as a resonance and cannot be 
observed as a free particle. 

Since the existence of the novel meson states is due to 
the Pauli principle it is immediately clear that the above  
confinement mechanism works only for quarks but 
not for gluons. But the confinement of gluons could possibly 
be a `second order effect' in Gribov's scenario, namely 
due to their coupling to light quarks which subsequently 
decay as described above. 

It is obviously desirable to find a quantitative description 
for this interesting physical picture of supercritical color 
charges. The confinement of quarks and gluons should 
be encoded in the singularities of the respective Green functions. 
Therefore the Green function of the quark is a suitable 
object to study in this context. 
In \cite{Lund,EPJ1} Gribov derived an equation 
for the retarded Green function of light quarks. It takes into 
account especially the dynamics of the infrared 
region but also reproduces asymptotic freedom at large 
momenta. Chiral symmetry breaking 
has been found to occur when the strong coupling constant 
exceeds a critical value, leading to the emergence 
of Goldstone boson (pions). It has been argued in 
\cite{EPJ1} that the nature of these Goldstone bosons is such 
that they should in fact be regarded as elementary objects. 
Corrections to the Green function caused by these 
Goldstone bosons are expected \cite{EPJ2} to lead a Green 
function of light quark which exhibits confinement, 
whereas the equation without these corrections is not expected 
to imply confinement \cite{Lund,EPJ1}. Unfortunately, 
the paper \cite{EPJ2} remained unfinished, and a full 
study of the analytic properties of the Green function 
and their consequences still 
remains to be done. Especially, it will be important to 
see how and to what extent the --- though somewhat 
simplified --- physical picture described above can be derived 
from the analytic properties of the resulting Green function. 

Gribov's equations for the Green function of light quarks 
(with or without pion corrections) 
are non--linear differential equations. 
So far the studies of these equations \cite{Lund,EPJ1,EPJ2} 
have been performed only by means of asymptotic expansions. 
It is the purpose of the present paper to perform a complete 
numerical study of the equation without pion corrections. 
This allows us to study the breaking of chiral symmetry 
also quantitatively and in more detail. We also investigate the 
analytic structure of the Green function resulting from 
Gribov's equation. 

It turns out that in Gribov's equation the critical value of the strong 
coupling constant is surprisingly low, $\alpha_c=0.43$. 
In the derivation of the equation it is assumed that the coupling 
constant does not become very much larger than this critical 
value. To some extent Gribov's approach can thus be 
considered as a semi--perturbative approach to confinement. 
This picture might also explain why we observe an 
essentially smooth behaviour of non--perturbative effects 
in the transition from the parton level to the hadron level. 
Typical multiplicities at the parton level, for example, are 
in surprising correspondence to those observed at the hadron 
level (for a more detailed discussion of this and similar 
observations see \cite{YuriVancouver}). 
The idea of an infrared finite coupling has also been widely discussed 
in the context of power corrections and the dispersive approach 
to renormalons in QCD \cite{BPY,power}, see also \cite{Martin} 
and references therein. In that approach, 
it appears consistent to define an effective running coupling down 
to very low momentum scales, in the sense that its first moments 
have a universal meaning. 
The values of the coupling found in the corresponding experimental 
analyses are in fact bigger than the critical value in Gribov's equation. 

The equation can be derived from the Dyson--Schwinger 
equation for the Green function of the quark in Feynman gauge. 
The approximations made in the derivation are motivated 
by the underlying physical picture, especially concerning the 
behaviour of the strong coupling constant. This method can 
therefore also be viewed as an unconventional approach 
to the difficult problem of solving the Dyson--Schwinger 
equations in QCD (for a review see \cite{DSreview}). 
The approximations usually made in solving 
the Dyson--Schwinger equations are intrinsically difficult 
to control. Comparisons with results obtained in Gribov's 
approach will therefore be potentially very useful.  

The paper is organized as follows. In section \ref{eqsection} we 
outline the main steps leading to the equation for the Green 
function of light quarks and describe some of its most 
important properties. In section \ref{param} a suitable 
parametrization of the Green function is given. The 
asymptotic behaviour of the equation for small and large 
momenta is discussed and the critical value of the strong 
coupling constant is derived. Section \ref{ImEuklidischen} 
deals with the Euclidean region of space--like momenta. 
In section \ref{asymfree} the solutions are shown to exhibit 
asymptotic freedom at large space--like momenta. 
Section \ref{models} provides models for the running 
coupling at small (space--like) momenta which are 
needed for the numerical analysis of the equation. 
The possible types of solutions in the Euclidean region 
are classified in section \ref{solutionseucreg}. The 
characteristic change in the solutions at supercritical 
coupling is discussed. Section \ref{secmassren} deals 
with the behaviour of the dynamical mass function of the 
quark in the Euclidean region. Phase transitions are 
found to occur for supercritical coupling and lead to 
chiral symmetry breaking. We study how this effect 
depends on the models used for the running coupling 
in the infrared. In section \ref{analyticstruct} 
we determine the analytic structure of the solutions 
in the whole momentum plane for the different 
types of solutions classified in section \ref{solutionseucreg}. 
We close with a summary and an outlook. 

The results presented in sections \ref{asymptbe} and 
\ref{asymfree} concerning the asymptotic behaviour 
of the equation have partly been obtained already in 
\cite{Lund}. They have been included in some detail in 
the present paper since they are immediately relevant 
to our analysis and serve to make it self--contained. 

\section{The equation for the Green function of light quarks}
\label{eqsection}

In this section we will outline the main steps that lead to the 
equation for the Green function of light quarks 
and and highlight some of its properties 
which are relevant to our discussion. 
Some of these properties and the full derivation of the equation 
have been discussed in detail in \cite{Lund,EPJ1}. 

\begin{figure}[htbp]
\begin{center}
\input{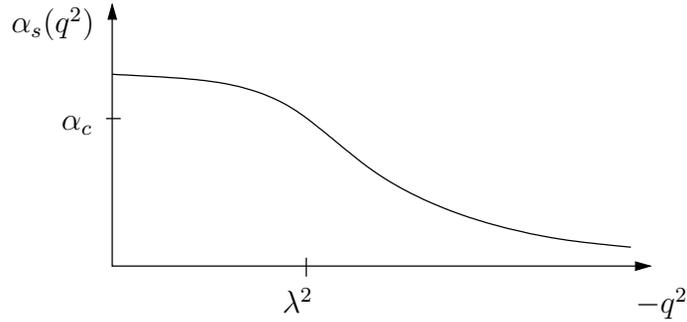}
\end{center}
\caption{Assumed behaviour of the strong coupling $\alpha_s(q^2)$}
\label{fig:Kopplung}
\end{figure}

The first step is the choice of a gauge. As noted by Gribov, 
the Feynman gauge turns out to be particularly well suited 
for deriving a simple  equation which is especially sensitive 
to the infrared dynamics. In other gauges it would be 
extremely difficult to find a similarly simple equation. The 
physical results, like for example the occurrence of chiral 
symmetry breaking, will of course be independent of the 
choice of gauge. 
In Feynman gauge the gluon propagator has the form 
\be
\label{effphot}
D_{\mu\nu}(k) = - \fr{g_{\mu\nu}}{k^2} \,\alpha_s(k^2)
\,. 
\ee
The exact behaviour of the strong coupling constant $\alpha_s$ 
is not known at small momenta. In Gribov's derivation of the 
equation it is assumed that the coupling constant is a slowly 
varying function of the momentum and does not become very 
large at small momenta. Such a behaviour is sketched in 
Fig.\ \ref{fig:Kopplung}. 
It turns out that the occurrence of a supercritical behaviour of the 
Green function does not depend on the details of the coupling 
in the infrared, as long as its value is above the critical value in 
some interval of momenta. As we will see, this critical value 
is rather low, $\alpha_c = 0.43$. 
These properties of the running coupling are consistent 
with the picture arising in the dispersive approach \cite{BPY} 
to power corrections 
in QCD (for reviews see \cite{YuriVancouver,Martin}). 
There it appears 
that the definition of a running coupling constant at very 
low momenta is possible in the sense that its integral moments 
have a universal meaning. Motivated by this, possible models 
for the coupling have been constructed, see for example 
\cite{Bryanalpha,Shirkov}. For our numerical study we will 
choose a rather simple form of the coupling, see section 
\ref{models} below. 

One starts from the Dyson--Schwinger equation for the inverse 
Green function $G^{-1}$ of the quark and considers its perturbative 
or diagrammatic expansion. To the corresponding sum of diagrams 
one applies the double 
differentiation $\del^2 =\del^\mu \del_\mu$, where $\del_\mu$ 
is the derivative with respect to the external momentum $q^\mu$ 
of the quark. Firstly, this is a way to regularize the divergences in 
these diagrams, and gives a finite result. 
Secondly, it can be used to collect the most singular contributions 
to the quark Green function from the infrared region. This is based 
on the observation that the action of $\del^2$ on the gluon 
propagator in Feynman gauge gives a delta function, 
\be
\del^2 \fr{1}{(q-q')^2 +i \epsilon} = - 4 \pi^2 i \,\delta^{(4)}(q-q') 
\,.
  \label{deltafunktion}
\ee
The integration variables in all diagrams can be arranged in such 
a way that the external momentum of the quark is carried along gluon 
lines. If a gluon line is now differentiated twice, 
the above identity then transforms the integral over the 
corresponding gluon momentum into two zero--momentum 
gluon insertions. 
All other contributions to the respective integrals 
--- those with derivatives in different lines or of the running 
coupling --- are clearly less singular in the infrared region. 
It is in principle possible to treat these terms systematically as 
corrections. But the resulting equation will then be a complicated 
integro--differential equation, as briefly indicated in 
\cite{YuriErice}. 
In first approximation those contributions are neglected, and 
one is left with a series of diagrams with two gluon insertions 
that carry zero momentum. This sum can 
be shown to be the diagrammatic expansion of a 
full inverse quark Green function with two full quark--gluon 
vertices $\Gamma_\mu(q,k=0)$ inserted. Using Ward 
identities the latter can be replaced by derivatives $\del_\mu G^{-1}$ 
of the quark Green function. 
Having eliminated the vertex functions, one ends up with 
a second order differential equation for the Green function 
of a light quark, 
\be
\del^2 G^{-1} = g (\del^\mu G^{-1}) \,G \,(\del_\mu G^{-1}) 
\,,
\label{gribgl}
\ee
where
\be
g = C_F \fr{\alpha_s(q)}{\pi} \,. \label{QCDdefg}
\ee
This is Gribov's equation which will be the subject of our study. 

A comment is in order concerning the choice of scale of 
the running coupling in (\ref{QCDdefg}). 
As described so far, the derivation of the equation has 
concentrated on the most important contributions from 
the infrared region. 
But it is of course desirable to find 
an equation which describes the Green function correctly 
also in the ultraviolet region. The use of the relation 
(\ref{deltafunktion}) implies that the coupling has to 
be evaluated at zero momentum. But it can be shown that 
by replacing $\alpha_s(0)$ by $\alpha_s(q)$ one 
arrives at an equation that also reproduces the correct 
behaviour at large momenta. Given the assumptions 
about the running coupling discussed earlier, the correction 
induced by this replacement is subleading as far as its 
contribution to the infrared region is concerned. 
In the approximation presently considered we can therefore 
accept equation (\ref{gribgl}) with (\ref{QCDdefg}) as 
an equation that is expected to provide an adequate description 
of  the Green function at all momentum scales. 

For simplicity, equation (\ref{gribgl}) is written for one--flavour 
QCD. This is sufficient as long as we are mainly interested 
in the occurrence of confinement and chiral symmetry breaking. 
The generalization to the more realistic case of a doublet 
of light quarks is straightforward and will be important for 
the study of bound states in Gribov's picture. 

The fact that equation (\ref{gribgl}) is a second order differential 
equation implies that its solutions will involve two dimensionful 
constants of integration. These will be related to the quark 
mass and the quark condensate. 

An obvious and important property of the equation is its 
invariance under a rescaling of the Green function, 
$G \to c G$ for any constant $c$. As a consequence, the 
equation will not involve the full wave function 
renormalization but only its logarithmic derivative. 
The equation is not scale invariant with respect 
to the momentum. The breaking of scale invariance 
is due to the presence of the running coupling constant. 
It is only through the running of the coupling that 
a momentum scale is introduced. 

A further property of the equation is a certain symmetry 
between the Green function $G$ and its inverse $G^{-1}$. 
One can easily show that the equation  (\ref{gribgl}) implies
\be
\del^2 G = (2 - g) (\del^\mu G) G^{-1} (\del_\mu G) \,.
\ee
This means that $G$ solves the same equation as $G^{-1}$, 
but with $g$ replaced by $2-g$, the symmetry point being 
$g=1$.
At $g=2$, corresponding to $\alpha_s=3 \pi/2$, the Green 
function would thus become a free Green function. 
This symmetry is certainly unphysical, and we should 
trust the equation only for comparatively small values 
of the coupling, roughly speaking below one or two. 
This is in agreement with the fact that 
in the derivation of the equation the coupling was assumed 
to be small. 

\section{Parametrization and asymptotic behaviour}
\label{param}

\subsection{Parametrization}

Due to invariance under parity and Lorentz transformations 
the inverse Green function has the general form 
\be
\label{ab}
G^{-1}(q) = a(q^2) \qs - b(q^2)    
\ee
with two scalar functions $a$ and $b$. 
We will in the following use the variable 
\be
\label{defofq}
q \equiv \sqrt{q^\mu q_\mu} \,,
\ee
such that the half plane $\real q \ge 0$ already covers 
the full plane in $q^2$, the variable in which the Green 
function is usually discussed. 
It will be convenient to use instead of (\ref{ab}) the following 
parametrization of the Green function\footnote{This 
parametrization deviates from the one used in 
\protect\cite{EPJ1,EPJ2}.}, 
\be
\label{Parametrisierung}
G^{-1} = - \rho \,\exp\lbr- \frac{1}{2} \phi \frac{\qs}{q}\rbr 
\ee
with two complex functions $\rho$ and $\phi$. 
This corresponds to 
\bea
a(q^2) &=& \frac{1}{q}\,\rho \,\sinh \frac{\phi}{2} 
\label{Parama} \\
b(q^2) &=& \rho \,\cosh \frac{\phi}{2} 
\eea
in the parametrization (\ref{ab}). 
The dynamical mass function $M$ of the quark is then given 
by the function $\phi$ only, 
\be
\label{Massenfkt}
M(q^2) = \frac{b(q^2)}{a(q^2)} = q\,\coth \frac{\phi}{2}
\,,
\ee
whereas the function $\rho$ represents the wave function 
renormalization. In terms of the usual notation we have 
$Z^{-1}=\rho/q$. 
We further introduce 
\be
\xi \equiv \ln q = \ln \sqrt{q^\mu q_\mu}
\ee
and denote the derivative with respect to this variable as 
\be
\dot{f}(q) = \del_\xi f(q) 
\,,
\ee
Since the solutions of Gribov's equation (\ref{gribgl}) 
depend only on the logarithmic derivative of the wave 
function renormalization, it is useful to define 
\be
\label{defp}
p = 1 + \beta \fr{\dot{\rho}}{\rho}  
\,.
\ee
where 
\be
\beta = 1 - g = 1- C_F \fr{\alpha_s}{\pi} 
\,.
\ee
Then the equation (\ref{gribgl}) for the Green function 
translates into a pair of coupled differential equations 
for $p$ and $\phi$, 
\be
\label{pdot}
\dot p = 1 - p^2 - \beta^2 \lbr \frac{1}{4} \dot\phi^2 
+ 3 \sinh^2 \frac{\phi}{2} \rbr 
\ee
\be
\label{phidotdot}
\ddot\phi + 2 p\,\dot\phi - 3 \sinh\phi = 0  
\,,
\ee
which will be the basis of our further analysis. 

\subsection{Asymptotic behaviour}
\label{asymptbe}

We now study the asymptotic behaviour of the 
solutions of Gribov's equation. An important outcome 
of this study will be the determination of the critical 
coupling at which chiral symmetry breaking occurs. 

First we keep the coupling constant fixed. 
The running of the coupling can then be treated under 
the assumption that the asymptotic behaviour 
of the solutions depends smoothly on the coupling. 
This assumption will be justified by our numerical 
analysis further below. Since the equations (\ref{pdot}) 
and (\ref{phidotdot}) depend only on the logarithm of 
$q$ the equations are the same along all straight lines 
passing through the origin of the complex $q$-plane. 
The initial conditions, at $q=0$ for example,  do not exhibit 
this apparent symmetry such that the solutions will be 
different in different directions in the $q$-plane. 
The fixed points of the equation, however, turn out to 
be independent of the direction in the $q$-plane. 

\bmat
\subsubsection*{Behaviour for $|q|\!\rightarrow \! \infty$}
\ubmat

For large $|q|$ the pair of equations (\ref{pdot}), 
(\ref{phidotdot}) has stable fixed points at 
\be
\label{phifix} 
\phi = (2 n+1) i \pi\,\;\;\;(n \in \Z) \,; \quad \quad \quad
p = \sqrt{1+3 \beta^2} \,. 
\ee
As we will see in section \ref{asymfree} the 
existence of these fixed points implies the asymptotic 
freedom of the corresponding solutions. 

The periodicity of the above fixed points is obvious from the 
equations, and we will now concentrate on the fixed point at 
$\phi=i \pi$. 
Perturbing the solutions around the fixed point, 
\be
\phi=i\pi + \psi \,;\;\;\;\;\; p = p_0 + \hat{p}
\ee
and expanding to first order in the perturbations we find 
\be
\label{p_0^2}
p_0^2 = 1 + 3 \beta^2 > 0 \,. 
\ee
Further we have  
\be
\del_\xi \hat{p} = - 2 p_0 \hat{p} \,, 
\ee
such that $\hat{p} = D e^{ - 2 p_0 \xi}$ with some $D \in \C$. 
For a stable fixed point we thus have to choose the positive 
root $p_0 = \sqrt{1+3 \beta^2} >0$ in (\ref{p_0^2}). 
We note that the function $\rho$ consequently develops a 
singularity
\be
\rho \sim \exp \lbk \fr{p_0 - 1}{\beta}\, \xi \rbk  \,. 
\ee
The linearized equation for $\psi$ becomes 
\be
\label{psilin}
\ddot{\psi} + 2 p_0 \dot{\psi} + 3 \psi = 0\,. 
\ee
Thus $\psi = C_1 e^{\gam_+ \xi} + C_2 e^{\gam_- \xi}$ 
with $C_1, C_2 \in \C$. We find 
\be
\label{gampm}
\gam_\pm = - p_0 \pm \sqrt{3 \beta^2 -2} 
\,.
\ee
Here $\gam_+$ and $\gam_-$ can be real  ($\beta^2>2/3$ ) 
or complex ($\beta^2<2/3$). As we will see in 
section \ref{secmassren} these two possible 
cases have quite different physical consequences. 

In the first case, $\beta^2 > 2/3$, the function $\phi$ 
approaches $i \pi$ monotonically. This case is characterized by 
\be
g < g_c = 1- \sqrt{\fr{2}{3}} \simeq 0.18
\ee
or 
\be
\label{gwiederunterkrit}
g > 1 + \sqrt{\fr{2}{3}} \simeq 1.82 \,.  
\ee
Here we find the critical value of the coupling constant $\alpha_s$ 
already mentioned earlier, 
\be
\alpha_c = \fr{3 \pi}{4} g_c \simeq 0.43  \,,
\ee
at which the solutions change their behaviour. 

In the second case, $\beta^2<2/3$, the function $\phi$ 
oscillates while approaching the fixed point $i \pi$. 
The corresponding supercritical behaviour of the 
equation is characterized by values of the strong coupling 
constant $\alpha_s$ in the interval 
\be
\alpha_c < \alpha_s < 4.3
\,.
\ee
The emergence of the upper limit 
is in agreement with the symmetry of 
the equation relating the Green function to its inverse 
for $g \to 2-g$, see section \ref{eqsection}. 
The fact that the equation exhibits subcritical behaviour 
at very large values of the coupling is certainly unphysical. 
We cannot expect that the equation describes the Green function 
correctly also at very large coupling. 

\bmat
\subsubsection*{Behaviour for $|q|\!\rightarrow \! 0$}
\ubmat

For small $|q|$ there are two possible cases. In the first case, 
$\phi$ approaches one of the fixed points described above, 
$\phi = (2 n+1) i \pi$. To show this one can proceed as 
in the case of large $|q|$. But now we are considering 
$\xi \rightarrow - \infty$ and therefore have to choose the 
negative root in (\ref{p_0^2}) in order to find a stable 
fixed point, $p_0 =- \sqrt{1+3 \beta^2}<0$. 
Again, the solutions will oscillate while approaching the 
fixed point if the coupling is supercritical. 

The other case possible for $|q|\rightarrow 0$ is 
that $\phi$ vanishes at $q=0$. 
Linearizing the equation for small $\phi$ results in  
\be
\label{linphiqklein}
\ddot{\phi} + 2 p_0 \dot{\phi} - 3 \phi = 0  \,.  
\ee
From the equation for $p$ we find that for this fixed point 
$p \rightarrow p_0$ with $p_0^2=1$. 
With the ansatz $\phi=C e^{\gam \xi}$ 
it is required that $\gam>0$ for $\phi$ to be regular. 
In order to have a solution which at large $q$ approaches 
$i \pi$ with damping we need $p_0=1$ and therefore $\gam=1$. 
The easiest way to see this is from eq.\ (\ref{glfuerepsilon}) 
and the corresponding discussion in section \ref{ImEuklidischen} 
below. 

\subsubsection*{Running coupling}

The running of the coupling can be treated 
assuming that the asymptotic behaviour of the solutions depends 
smoothly on it. The values of the function $p$ at the fixed points 
discussed above depend on $\beta$. Therefore they are changed 
accordingly, i.\,e.\ have to be replaced by $\beta(q=0)$ or 
$\beta(q=\infty)=1$, respectively. The oscillations occurring 
for $|q|\rightarrow \infty$ stop at the scale at which the coupling 
becomes smaller than $\alpha_c$, 
and $\phi$ approaches the fixed point monotonically above 
this scale. 

\section{Solutions in the Euclidean region}
\label{ImEuklidischen}

We first investigate the equation in the Euclidean region, 
i.\,e.\ for space--like momenta. 
Therefore we want to consider purely imaginary 
values of our variable $q$, thus $q=i \qt$ with a 
real--valued and positive $\qt$. 
The derivative with respect to $\qt$ will be denoted 
\be
\label{defqprime}
\frac{d}{d\qt} f= f'(q)
\,.
\ee
For space--like momenta the dynamical mass function 
$M(q^2)$ is required to be real--valued. 
Eq.\ (\ref{Massenfkt}) then implies that $\phi$ is purely 
imaginary. In addition, a real--valued 
mass function requires that the function $p$ is real--valued 
for space--like momenta. 
For convenience we define for the use in the 
present section 
\be
\phi = i \chi \,,
\ee
where $\chi$ is real--valued. In this section we thus have to 
consider only real--valued functions $\chi$ and $p$ depending 
on the real parameter $\qt$. 

The equation (\ref{phidotdot}) for $\chi$ (or $\phi$, respectively) 
can be reformulated in such a way that it permits a simple 
interpretation. The function 
\be
\label{defepsilon}
\epsilon \equiv \fr{\dot{\chi}^2}{2} - 3 \,(1- \cos\chi) 
\ee
can be interpreted as the energy of a motion with 
$\chi$ being a one--dimensional degree of freedom. 
The equation of motion equivalent to (\ref{phidotdot}) is  
\be
\label{glfuerepsilon}
\del_\xi \epsilon =  - 2 p \dot{\chi}^2  \,.
\ee
The behaviour of $\chi$ can then be interpreted as a motion 
with damping (given by $p$) in the potential 
\be
\label{potentialV}
V = - 3 \,(1- \cos\chi)  \,.
\ee
This potential has minima at $\chi=(2n +1) \pi$ for all $n\in \Z$. 
Thus for space--like momenta the fixed points discussed in 
section \ref{asymptbe} appear as the minima of the potential $V$. 

\subsection{Asymptotic freedom}
\label{asymfree}

For large $Q^2=-q^2$ the Green function should behave 
according to perturbative renormalization and exhibit 
asymptotic freedom. We now show that Gribov's equation 
reproduces exactly this behaviour for solutions that approach 
one of the fixed points discussed above. This is done by 
considering the leading terms in the limit of large $\qt$. 

We first consider the wave function renormalization. The 
corresponding renormalization constant is in our parametrization 
defined as 
\be
\label{defwellenrenorm}
\rho = e^\xi Z^{-1}(\xi) = q Z^{-1} 
\,,
\ee
as can be seen when eq.\ (\ref{Parama}) is evaluated 
at $\phi\simeq i \pi$. As is usually done we assume $Z(\xi)$ to 
be a slowly varying function. 
Using (\ref{defp}) and (\ref{pdot}) one derives in the 
limit $\chi \to \pi$ the following 
equation\footnote{In \cite{Lund} this equation (there eq.\ (4.42)) 
contains a misprint.} for $\rho$, 
\be
\label{asympglrho}
(\del_\xi + 3)(\del_\xi - 1) \rho + 3 g \rho 
- g \fr{1}{\rho}(\del_\xi \rho)^2 = 0 
\,. 
\ee
Inserting (\ref{defwellenrenorm}) and neglecting 
terms of the order $\del^2_\xi Z^{-1}$ and $(\del_\xi Z^{-1})^2$ 
we find 
\be
\label{wellenrenorm}
\del_\xi Z^{-1} + \fr{1}{2} g Z^{-1} = 0 
\,. 
\ee
This coincides with the well--known wave function renormalization 
in Feynman gauge as it can be found for example in \cite{Buras}. 
Turning to mass renormalization we observe that 
close to one of the fixed points, $\phi= i \chi = i (\pi-\sigma)$ 
with small $\sigma$, the mass function is given by 
(see eq.\ (\ref{Massenfkt}))
\be
\label{mfuerrein}
\sigma = 2 e^{-\xi} M(\xi) = \fr{2 M}{\qt} 
\,.   
\ee
Linearizing eq.\ (\ref{phidotdot}) we find 
\be
\label{linsigmamass}
\ddot{\sigma} + 2 \lbr 1 
+ \beta \fr{\dot{\rho}}{\rho} \rbr \dot{\sigma} 
+ 3 \sigma = 0  
\,.
\ee
We further note that due to (\ref{defwellenrenorm}) and 
(\ref{wellenrenorm}) 
\be
\fr{\dot{\rho}}{\rho} = 1 + \fr{\del_\xi Z^{-1}}{Z^{-1}}
   = 1 - \fr{1}{2} g 
\,.
\ee
Together with (\ref{mfuerrein}) this can be inserted 
in (\ref{linsigmamass}). Neglecting the term of order 
$\del^2_\xi M$ we arrive at 
\be
\label{massenrenormgl}
\del_\xi M = - \fr{3}{2} g M 
\,,
\ee
which is exactly the mass renormalization at one loop. 
Its solution is 
\be
  \label{massenrenorm}
M(q^2) = m_0 \lbk \fr{\alpha_s(q^2)}{\alpha_s(q_0^2)} 
   \rbk^{\gam_m}
\,,
\ee
where $m_0$ is the mass at a given scale $q_0$, and in one--loop 
approximation the exponent is $\gam_m=4/b_0$ with 
$b_0=11-\frac{2}{3} n_f$. 

Taking into account also the sub--leading solution for $\sigma$ 
from (\ref{linsigmamass}) or, equivalently, from (\ref{psilin}) we 
find that $\sigma$ behaves at large $\qt$ as 
\be
\label{sigmacondensate}
\sigma \sim \frac{2 m_q}{\qt} + \frac{\nu^3}{\qt^3} 
\,,
\ee
as can be seen from (\ref{gampm}) since in this limit $\beta \to 1$. 
Accordingly, the mass function behaves as 
\be
M(q^2) \sim m_q -\frac{\nu^3}{\qt^2}
\,.
\ee
We thus find two dimensionful parameters, $m_q$ and $\nu^3$. 
These can be  identified with the quark mass 
and a quark condensate, respectively. 
It is difficult to disentangle the two terms in (\ref{sigmacondensate}) 
numerically. We will therefore not pursue this interesting issue 
any further in the present paper. 

\subsection{Models for the running coupling}
\label{models}

The behaviour of the strong coupling constant is 
only very vaguely known in the infrared region. 
In order to perform a numerical study 
we have to use a model for $\alpha_s$. 
The model is required to be in agreement 
with the general assumptions used in the derivation 
of the equation (see section \ref{eqsection}). 
Obviously, any useful model should coincide 
with the perturbative running of the coupling at large 
momentum scales. In the following we will use 
two different models of this kind. Both are rather simple 
but should be sufficient for studying the physical effects 
resulting from Gribov's equation. More complicated 
models (see for example \cite{Bryanalpha,Shirkov}) could 
of course be implemented in the same way. 

For the perturbative behaviour of the coupling we have to specify 
$\Lambda_{\mbox{\tiny QCD}}$ and $n_f$. 
Moderate changes in these parameters do not have any 
significant effect on our results for the Green function since 
they only change the behaviour of $\alpha_s$ at large 
momenta where the coupling is subcritical. To be specific 
we choose $\Lambda_{\mbox{\tiny QCD}} =  250\,\mbox{MeV}$ 
and $n_f=3$. (The latter choice is not completely in agreement 
with the fact that the equation is in the present paper studied for only 
one flavour. The choice $n_f=1$ would lead to almost identical 
results for the Green function.) 

\begin{figure}[htbp]
\begin{center}
\input{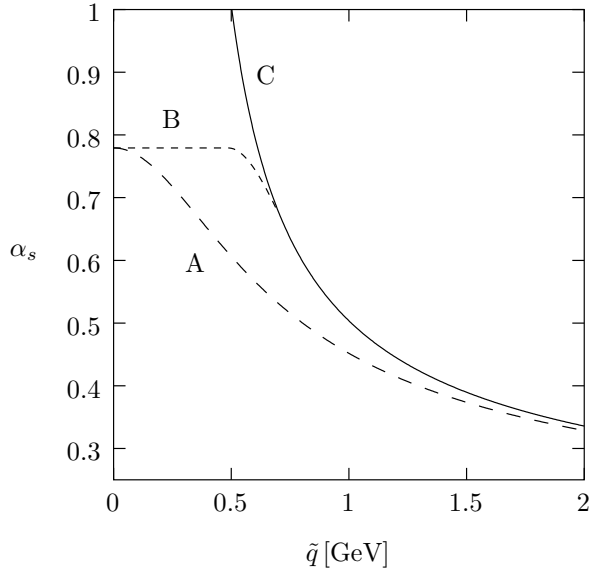}
\end{center}
\caption{Models for the strong coupling 
constant (A, B) and the behaviour according 
to one--loop renormalization (C)}
  \label{fig:alpha}
\end{figure}

Our first model is the more realistic one, and is motivated by the 
dispersive approach to power corrections in QCD \cite{BPY}. 
This approach is based on the assumption that the coupling constant 
can be defined down to very small momenta, and that this coupling 
in the infrared has a universal meaning. Then it is possible to determine 
its integral over the infrared region from measurements 
of infrared and collinear safe observables like certain event shape 
variables (for a recent review see \cite{YuriVancouver}). 
In this way one finds for the integral of the coupling 
\be
\label{intofcoupling}
\alpha_0= 
\frac{1}{2\,\mbox{GeV}} 
\int_0^{2\,\mbox{\scriptsize GeV}} \!\alpha_s(k) \, dk
\simeq 0.5 
\,.
\ee
This condition can be fulfilled by shifting the argument 
of the logarithm in the usual one--loop formula for the 
perturbative running of the coupling, 
\be
\label{Kopplungma}
\alpha_s(q^2)  = \fr{4 \pi}{\lbr 11-\fr{2}{3} n_f \rbr \, \ln  
  ( -q^2 / \Lambda^2_{\mbox{\tiny QCD}} +a)} 
\,. 
\ee
For $a=0$ this is exactly the one--loop renormalization 
of the coupling. Our first model for the running 
coupling is obtained for $a=6$. This choice is made to satisfy 
the condition (\ref{intofcoupling}). 
We will in the following refer to this model as type A. 
This running of the 
coupling is shown as curve A in Fig.\ \ref{fig:alpha}. 
For comparison we show in that figure as curve C also the 
coupling obtained from the one--loop renormalization. 

The second model is shown as curve B in Fig.\ \ref{fig:alpha}. 
It is obtained from the one--loop renormalization of the 
coupling (see eq.\ (\ref{Kopplungma}) with $a=0$) by 
simply cutting it off at some given value and assuming it 
to be constant below the corresponding momentum scale.  
In order to avoid problems in the numerical treatment of 
the equation we smooth out the resulting edge by fitting 
a polynomial of third degree such that the first derivative 
is continuous. Apart from this detail the coupling in this 
model, which we will refer to as type B, is uniquely determined 
by a given value $\alpha_s(0)$ at vanishing momentum. 
This model is in general, i.\,e.\ for arbitrary $\alpha_s(0)$, 
not in agreement with the condition (\ref{intofcoupling}). 
It is not constructed to serve as a realistic description of the strong 
coupling.  Instead, it will mainly be used in order to study the 
qualitative effects of Gribov's equation. For this purpose 
a model is useful in which the coupling is clearly 
supercritical in a large region of momenta. 
It will also be useful to vary the strength of the coupling 
and to study the effects resulting from this change. 
Similarly, a comparison between the results obtained 
with coupling of type A and of type B will be interesting. 
The differences will in that case also depend on the 
initial conditions of the solutions. 

\subsection{The solutions in the Euclidean region}
\label{solutionseucreg}

We now turn to the numerical study of the pair 
of differential equations (\ref{pdot}), (\ref{phidotdot}) in 
the Euclidean region of space--like momenta. 
We have used two different numerical 
methods. All solutions presented below have been found 
using a Runge--Kutta procedure, 
i.\,e.\ the step--wise integration of the equation starting 
from a set of given initial conditions. 
We have also used the routine {\tt COLSYS} \cite{colsys}, 
a non--local collocation procedure using B-splines which is 
also suited for boundary--value problems with boundary 
conditions given at different points. Within numerical errors 
agreement has been found in all cases in which both methods 
have been applied. 

The solutions in the Euclidean region can be classified according 
to their behaviour at small and large momenta $\qt$. 
As discussed in section \ref{asymptbe}, the possible fixed points 
for $\qt \to 0$ and $\qt \to \infty$ differ from each other 
by the values of the function $\chi$ (or equivalently $\phi$). 
The values of $p$ are then fixed in the respective limits. 
Due to the $2 \pi$-periodicity of the equations in $\chi$ we can 
restrict ourselves in the following to the case in which 
$\chi \to \pi$ for $\qt \to \infty$. We will distinguish three 
classes of solutions in which the function $\chi$ approaches 
in the limit $\qt \to 0$ the values $0$ (or $2 \pi$), $-\pi$, 
or $\pi$, respectively. We will now discuss these three classes 
separately. 

\begin{figure}[htbp]
\begin{center}
\input{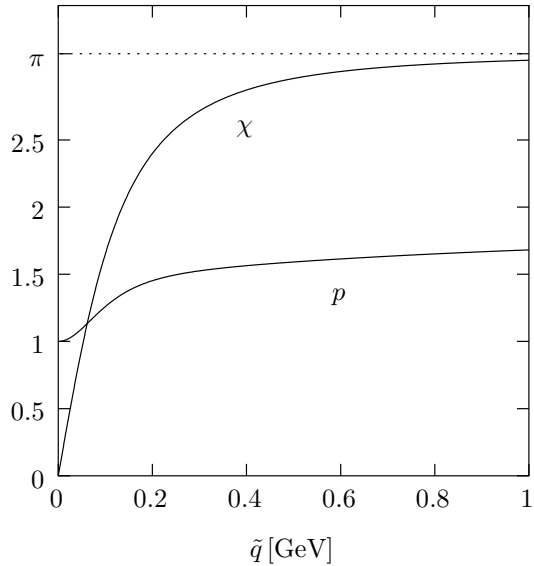}
\end{center}
\caption{Solutions $\chi$ and $p$ for $\chi'(0)\!=\!20$ 
(corresponding to $m_R\!=\!100$ MeV, see below) with 
running coupling of type A}
\label{fig:pert}
\end{figure}

\begin{figure}[htbp]
\begin{center}
\input{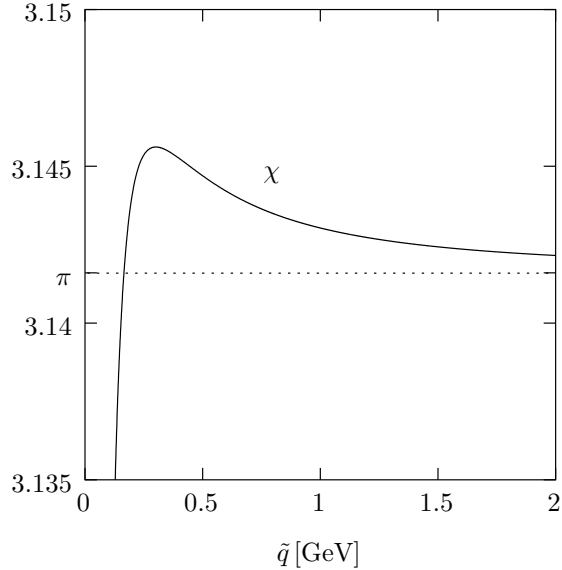}
\end{center}
\caption{Solution $\chi$ for $\chi'(0)\!=\!250$ 
(corresponding to $m_R\!=\!8$ MeV, see below) with 
running coupling of type A}
\label{fig:pertschwing}
\end{figure}

The solutions in the first class\footnote{These are the (only) 
solutions discussed in \protect\cite{Lund}.} start at 
$\chi(\qt=0) = 0$. 
As discussed in section \ref{asymptbe} this implies $p(0)=1$. 
The free parameter in this class of solutions is therefore 
$\chi'(0)$ (or, equivalently, the renormalized mass $m_R$, 
see section \ref{secmassren} below). 
Here and in the following we will measure $\qt$ in units 
of $\mbox{GeV}$, and thus $\chi'$ in units of $\mbox{GeV}^{-1}$. 
For small $\chi'(0)$ the function $\chi$ approaches $\pi$ 
monotonically. Such a solution is shown in Fig.\ \ref{fig:pert}, 
and we have plotted both $\chi$ and $p$. 
If the coupling constant is subcritical for all $\qt$ (for example 
for model B with $\alpha_s(0) < \alpha_c$) the solution 
is monotonic for all possible $\chi'(0)$. 
This situation changes if the coupling constant is supercritical 
in some interval of momenta. For small values of $\chi'(0)$ 
the solution is still monotonic, c.\,f.\ the solution in 
Fig.\ \ref{fig:pert} which is found for supercritical coupling. 
Due to the smallness of $\chi'(0)$ these solutions come close to 
$\pi$ only at momentum scales at which the running coupling 
is already subcritical again, and we do not observe oscillations. 
But for larger values of $\chi'(0)$ the function $\chi$ increases 
more rapidly and can pass the value $\pi$. Such a solution is shown 
in Fig.\ \ref{fig:pertschwing}. 
For even larger $\chi'(0)$ the function $\chi$ can pass $\pi$ 
more often, and in fact even arbitrarily often 
for $\chi'(0) \to \infty$. 
However, these oscillations are very strongly damped 
as the scale on the vertical axis in Fig.\ \ref{fig:pertschwing} 
illustrates. In all cases the solutions stop oscillating 
as $\qt$ increases. This was to be expected since the 
coupling becomes subcritical at larger momentum scales. 
If $\chi'(0)$ is negative the function $\chi$ approaches 
$-\pi$ instead of $\pi$ at large $\qt$. Due to the periodicity 
of the equations in $\chi$ we can then assume that 
these solutions start at $\chi(0) = 2 \pi$ (instead of $0$) and 
approach the fixed point $\pi$. Otherwise the behaviour of these 
solutions (oscillations, damping etc.) is not different from 
the case of positive $\chi'(0)$. 

\begin{figure}[htbp]
\begin{center}
\input{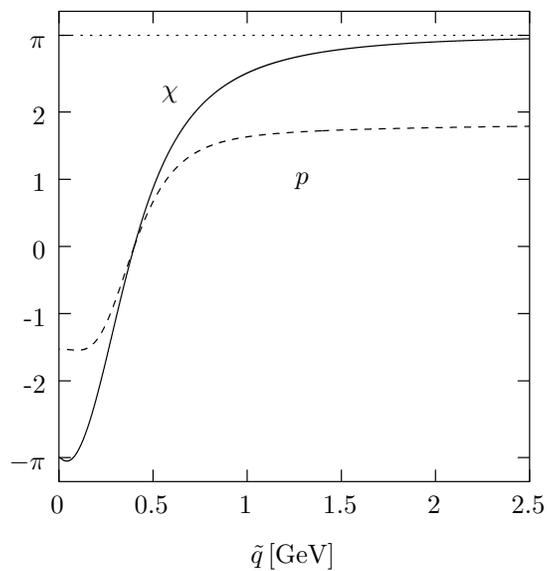}
\end{center}
\caption{Solutions 
$\chi$ and $p$ for $\chi(0.4)\!=\!0$, $\chi'(0.4)\!=\!10$ 
and $p(0.4)\!=\!0$ with running coupling of type A}
\label{fig:adrueber}
\end{figure}

In the second class of solutions $\chi$ runs from $-\pi$ to 
$\pi$ as $\qt$ runs from $0$ to $\infty$. In other words, 
$\chi$ goes from 
one minimum of the potential to a neighboring one. 
At the same time $p$ runs from $-\sqrt{1+3 \beta^2(0)}$ 
to $\sqrt{1+3 \beta^2(\infty)}=2$. It is possible 
that the zeros of $\chi$ and $p$ coincide. 
Such a solution is presented in Fig.\ \ref{fig:adrueber}. 
In a sense, these solutions are `symmetric'. (This term would 
be more appropriate if the coupling was not running but constant.) 
There are also solutions in which the zeros of $\chi$ and $p$ 
do not coincide, they are `asymmetric' in this sense. 

\begin{figure}[htbp]
\begin{center}
\input{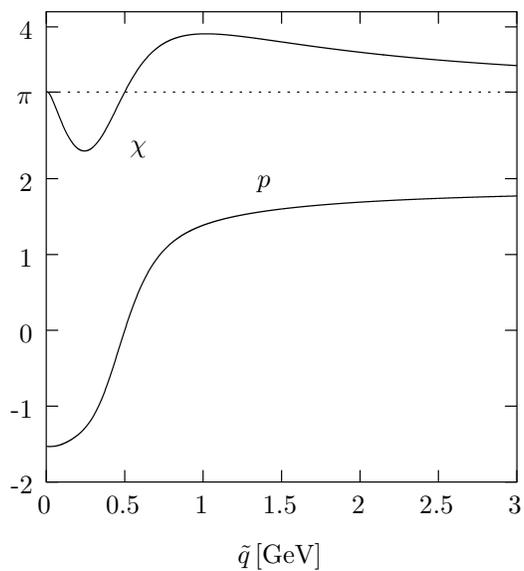}
\end{center}
\caption{Solutions $\chi$ and $p$ for $\chi(0.5)\!=\!\pi$, 
$\chi'(0.5)\!=\!4$ and $p(0.5)\!=\!0$ with 
running coupling of type A}
\label{fig:adrin}
\end{figure}

The third class comprises such solutions in which $\chi$ 
starts at $\pi$ and also approaches this value at large $\qt$. 
The solutions thus stay in one well of the potential. 
At the same time $p$ runs over the same range as in 
the previous class of solutions. Similar to that case, it is possible 
that $\chi=\pi$ and $p=0$ coincide. Such a `symmetric' 
solution is shown in Fig.\ \ref{fig:adrin}. 
Of course, there are also `asymmetric' solutions in which 
these special values of $\chi$ and $p$ occur at different momentum 
scales. 

If the coupling is supercritical the solutions of the second 
and third class can exhibit oscillations 
of $\chi$ around $\pi$ (and also around $-\pi$ for $\qt \to 0$ 
in the third class) 
similar to the ones described above for the first class. 

Numerically, the solution of the first class can 
be found by integrating the equation starting at $\qt=0$. 
In order to integrate the solutions numerically for the 
second and third class one has to start at some intermediate 
$\qt_0$. This is no restriction since the matching of two solutions 
for $0<\qt<\qt_0$ and $\qt_0<\qt<\infty$ is trivial. 
It just reflects the fact that integrating a differential equation 
starting from a stable fixed point leads to exponentially 
large numerical errors. 

The three classes discussed above exhaust all asymptotically free 
solution in the Euclidean region, i.\,e.\ solutions which end up in 
one of the fixed points for $\qt \to \infty$. 
Especially, there are no solutions in which $\chi$ runs from a 
minimum of the potential (\ref{potentialV}) to any other minimum 
that is not a neighboring one. 

\section{Mass renormalization and chiral symmetry breaking}
\label{secmassren}

In this section we would like to address the question how the 
dynamical mass function $M(q^2)$ of the quark behaves 
in the Euclidean region for the solutions discussed in the 
preceding section. We will mainly concentrate 
on the first class of solutions and only briefly comment on the 
other two classes at the end of the section. 

Let us define\footnote{This term was introduced in 
\protect\cite{Lund} and we adopt it here. This mass is 
not meant to be renormalized as opposed to being a bare 
mass in the usual sense. The mass function of course describes the 
mass renormalization for all $q^2$. This `renormalization' 
is done down to $\qt=0$, having in mind a starting point at 
large $\qt$.}  
the `renormalized' mass $m_R$ as the limit 
of the mass function $M(q^2)$ as the momentum vanishes,
\be
\label{defmR}
m_R = \lim_{\qt \to 0} M(q^2)
\,.
\ee
Since in the first class of solutions $\chi(0)=0$ we can 
expand (\ref{Massenfkt}) to find that the renormalized 
mass becomes 
\be
\label{expandmR}
m_R=\frac{2}{\chi'(0)}
\,.
\ee
Since $\chi'(0)$ was just the free parameter specifying these 
solutions we can use $m_R$ instead to characterize them uniquely. 
Although the renormalized mass is the small--momentum limit 
of the dynamical mass function it would most probably be 
too simple to interpret it as a constituent mass of the quark. 

Further we want to define a `perturbative' mass $m_P$ 
of a given solution. It is supposed to reflect the behaviour 
of the perturbative tail of the mass function of this solution. 
This could be achieved by computing its asymptotic 
behaviour which is eventually described by eq.\ (\ref{massenrenorm}).  
For our purposes it turns out 
to be sufficient and more convenient for the numerical 
study to choose a definition which involves only a finite 
momentum scale. 
We therefore define the perturbative mass as the value of 
the mass function at the scale $\lambda$ at which the coupling 
becomes subcritical (see also Fig.\ \ref{fig:Kopplung}) and the 
perturbative behaviour is expected to set in, 
\be
\label{defpertmass}
m_P= M(\lambda^2) = \lambda \cot\fr{\chi(\lambda)}{2}
\,.
\ee
Obviously, the scale $\lambda$ in 
this definition depends on the model for the running 
coupling. In the models A and B discussed in section 
\ref{models}, and presumably in most other realistic models, 
the values of $\lambda$ are very similar. 
Any other choice of scale would lead 
to similar results as long as that scale is chosen larger 
than $\lambda$. 

\begin{figure}[htbp]
\begin{center}
\input{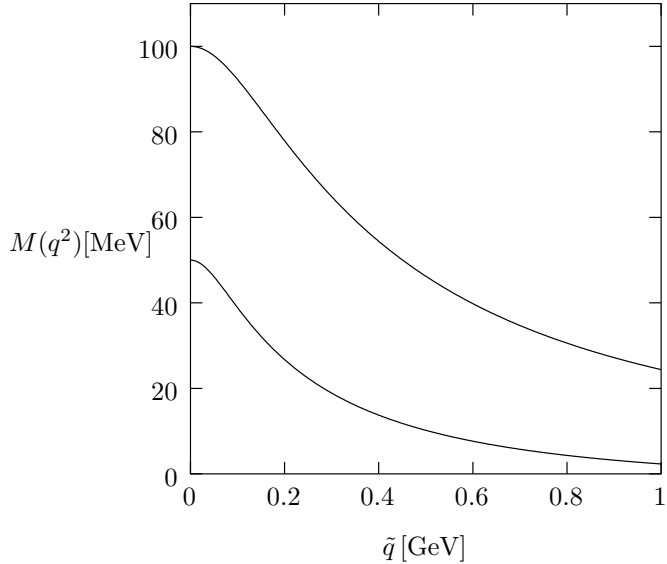}
\end{center}
\caption{The mass function $M(q^2)$ for 
   renormalized masses $m_R\!=\!50\,\mbox{MeV}$ and  
   $100 \, \mbox{MeV}$, respectively, for coupling of type A}
\label{fig:mabfall}
\end{figure}

A general property of the solutions of Gribov's equation 
is the rapid decrease of the mass function with increasing 
momentum. This is illustrated in Fig.\ \ref{fig:mabfall} 
for two solutions that are similar to the one shown in 
Fig.\ \ref{fig:pert}. 
Both solutions in this figure do not exhibit oscillations since 
they correspond to comparatively small values of $\chi'(0)$ 
(see also the corresponding discussion in section 
\ref{solutionseucreg}). 

\begin{figure}[htbp]
\begin{center}
\input{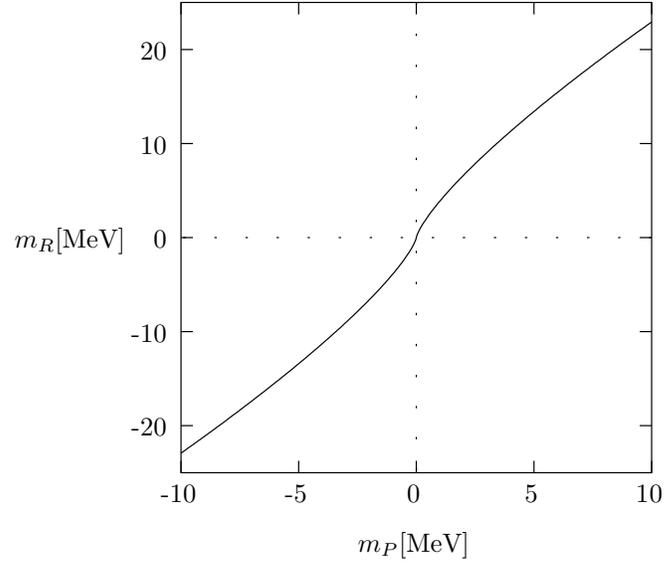}
\end{center}
\caption{Renormalized mass $m_R$ vs.\ 
 perturbative mass $m_P$ for subcritical coupling 
 (coupling of type B with $\alpha_s(0)\!=\!0.3$)}
\label{fig:m2munter}
\end{figure}

It is now interesting to study the relation between the 
renormalized mass $m_R$ and the perturbative mass $m_P$. 
First we consider the case of subcritical coupling. 
We choose model B for the running coupling with a 
maximal value $\alpha_s(0)\!=\!0.3$. Since the coupling 
is subcritical for all momenta the scale $\lambda$ cannot 
be defined via the critical value $\alpha_c$ in this model. 
Therefore we have to supplement 
the definition (\ref{defpertmass}) of the perturbative mass 
with the choice $\lambda= 1\,\mbox{GeV}$ in this case. This choice 
is to some extent arbitrary. It is mainly motivated by the values of 
$\lambda$ resulting from the models A or B with supercritical 
$\alpha_s(0)$. Other choices for $\lambda$ in the subcritical 
case lead to similar results for the dependence of $m_R$ 
on $m_P$. This dependence is shown in Fig.\ \ref{fig:m2munter}. 
For subcritical coupling there is a one--to--one correspondence 
between renormalized mass $m_R$ and perturbative mass $m_P$. 
The renormalized mass decreases with decreasing perturbative mass, 
and vanishes when the perturbative mass vanishes. Chiral symmetry 
is thus not broken at subcritical coupling. 

\begin{figure}[htbp]
\begin{center}
\input{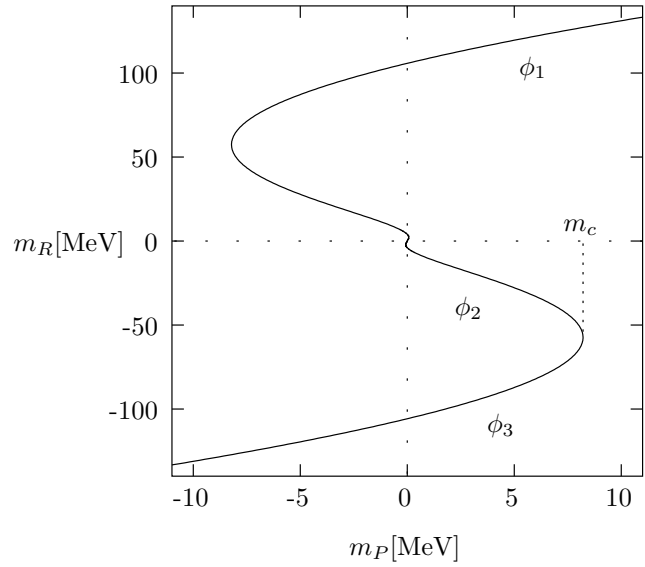}
\end{center}
\caption{Renormalized mass $m_R$ vs.\ 
  perturbative mass $m_P$ for 
  coupling of type B with $\alpha_s(0)\!=\!0.94$}
\label{fig:m2m}
\end{figure}

If the coupling is supercritical at small momenta some of the 
solutions exhibit oscillations. 
These are possible until the running coupling becomes subcritical, 
i.\,e.\ as long as $\qt<\lambda$. There are also monotonic 
solutions for which $m_R$ has to be sufficiently large 
(see section \ref{solutionseucreg}). Let us consider one of 
the latter. The corresponding function $\chi$ has a certain 
value at the scale $\lambda$. But there are also solutions $\chi$ 
with smaller $m_R=2/\chi'(0)$ which oscillate and pass $\pi$ twice. 
It is now possible that one of these solutions has the same value at 
the scale $\lambda$ as the monotonic solution considered before. 
There can in fact be even more solutions with this property, 
among them also solutions with negative $m_R$ (resp.\ negative 
$\chi'(0)$). The number of these solutions will (for $m_P \neq 0$, 
see below) remain finite. The reason for this is the following. 
A solution $\chi$ which oscillates very often will be very close to 
$\pi$ due to the strong damping in the corresponding equation. 
As a result there is a maximal possible $\chi(\lambda)$ which the 
solutions can reach for a given number of oscillations. 
In summary, we find that different values of $m_R$ (resp.\ 
$\chi'(0)$) can lead to identical values of $\chi(\lambda)$. 
But since the perturbative mass $m_P$ depends only 
on this value $\chi(\lambda$), see (\ref{defpertmass}), the 
correspondence between the renormalized and the perturbative 
mass is no longer one--to--one. Instead it takes the form 
shown in Fig.\ \ref{fig:m2m}, 
and the renormalized mass becomes a multi--valued 
function of the perturbative mass. 
Here we have chosen model B for the running coupling, thus 
$\lambda=1.27\,\mbox{GeV}$. Further we have 
chosen $\alpha_s(0)=0.94$ in this figure because the effect is 
more pronounced at large coupling. The dependence on the 
coupling strength and on the model will be studied below. 
The figure has been obtained by integrating the differential 
equation up to $\qt=\lambda$ with varying initial conditions 
$\chi'(0)$. 

Fig.\ \ref{fig:m2m} shows that at supercritical coupling 
the renormalized mass does not vanish in the limit of 
vanishing perturbative mass, and we thus find that 
chiral symmetry is broken. This figure is  
one of our main results and we will now study it in some 
more detail. 
At large perturbative mass there is only one branch 
of solutions, denoted as $\phi_1$ in the figure. 
But if the perturbative mass is below a critical 
mass $m_c$ there are two additional solutions with different 
renormalized mass, denoted as the branches 
$\phi_2$, $\phi_3$ in the figure, and we can regard 
this as a phase transition in the vacuum of light quarks. 
If we consider only 
the region of positive perturbative masses we can 
identify the branch $\phi_1$ with monotonic 
solutions, whereas the branch $\phi_3$ corresponds 
to solutions in which $\chi$ passes $\pi$. Further, 
the branch $\phi_2$ can be identified with solutions 
in which $\chi$ passes $\pi$ and in addition has 
a turning point. 

We observe that the curve in 
Fig.\ \ref{fig:m2m} is symmetric with respect 
to the origin. This is an immediate consequence of the fact 
that the equations (\ref{pdot}), (\ref{phidotdot}) are 
invariant under the exchange 
\be
\phi \rightarrow 2 \pi i - \phi
\,.
\ee
The mass function changes sign under this transformation, 
and hence the symmetry. 

\begin{figure}[htbp]
\begin{center}
\input{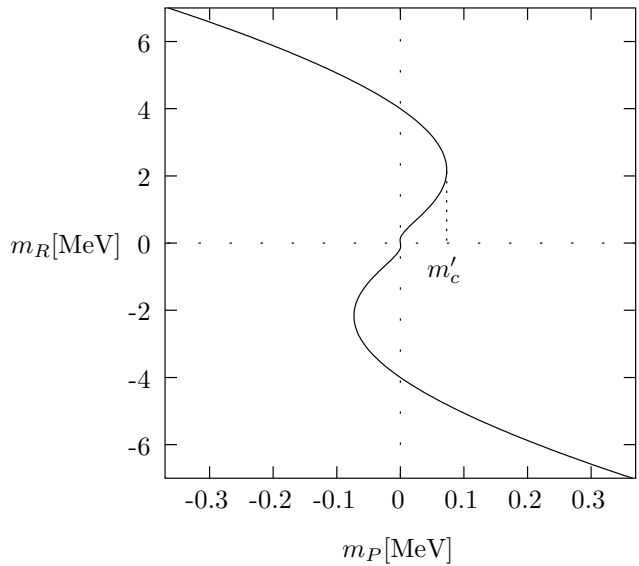}
\end{center}
\caption{Renormalized mass $m_R$ vs.\  
  perturbative mass $m_P$ for 
  coupling of type B with $\alpha_s(0)\!=\!0.94$, 
  detail of Fig.\ \protect\ref{fig:m2m}}
\label{fig:ausschn}
\end{figure}

The phase transition at the critical value $m_c$ of 
the perturbative mass $m_P$ 
is not the only one. In fact there is an infinite series 
of similar phase transitions taking place in the limit 
$m_P \to 0$. The curve in Fig.\ \ref{fig:m2m} exhibits 
a very interesting self--similarity which illustrates this 
series of phase transitions. In Fig.\ \ref{fig:ausschn} 
we show a detail of Fig.\ \ref{fig:m2m} around the 
origin. 
Both figures show different parts of the same curve. 
We denote the critical mass of the second phase transition 
as $m_c'$. In the two additional solutions occurring at 
this scale the function $\chi$ passes the value $\pi$ twice. 

It is interesting to note that in each phase transition 
one of the two additional branches of solutions has the quite 
unusual property that the renormalized mass grows 
with decreasing perturbative mass. 

The phase transitions lead to the generation of pions as Goldstone 
bosons \cite{Lund,EPJ1}. The Bethe--Salpeter amplitude 
of the pion can be shown to be $\varphi= C \{ G^{-1},\gamma_5 \}$ 
with a constant $C$. It solves an equation for $q\bar{q}$ bound 
states which is derived in a similar approximation scheme as 
the equation for the Green function \cite{Lund}.  
It has been speculated that the observation of just one pion 
in nature should restrict the physical value of the perturbative 
mass of the quark to be between the first and second 
phase transition \cite{Bass}. As can already be seen from Figures 
\ref{fig:m2m} and \ref{fig:ausschn}, 
however, two successive phase transitions happen to take place 
at mass scales which differ from each other by two 
orders of magnitude. Therefore it is not 
possible to deduce any considerable restriction on the physical 
perturbative mass of the quark in this way. 

It is now instructive to study the dependence 
of the critical mass scales on the value and on the model 
chosen for the running coupling. For this we use the 
two models A and B for the running coupling introduced in 
section \ref{models}. There we did not yet define how to 
vary the coupling strength in model A. We  
will use the value $\alpha_s(0)$ at vanishing momentum 
as a parameter (as in model B) and adjust the value of 
$a$ in eq.\ (\ref{Kopplungma}) accordingly to obtain 
a running coupling of varying strength in this model as well. 
Of course, the condition (\ref{intofcoupling}) will then 
no longer be fulfilled. 
In model B the momentum scale $\lambda$ 
at which the coupling becomes 
critical is independent of the value $\alpha_s(0)$, 
but in model A it depends by construction 
on the parameter $a$. Since this dependence is rather 
weak and in order to make the results comparable we use 
also for model A the fixed value 
$\lambda=1.27\,\mbox{GeV}$  which is obtained for 
model B. 

\begin{figure}[htbp]
\begin{center}
\input{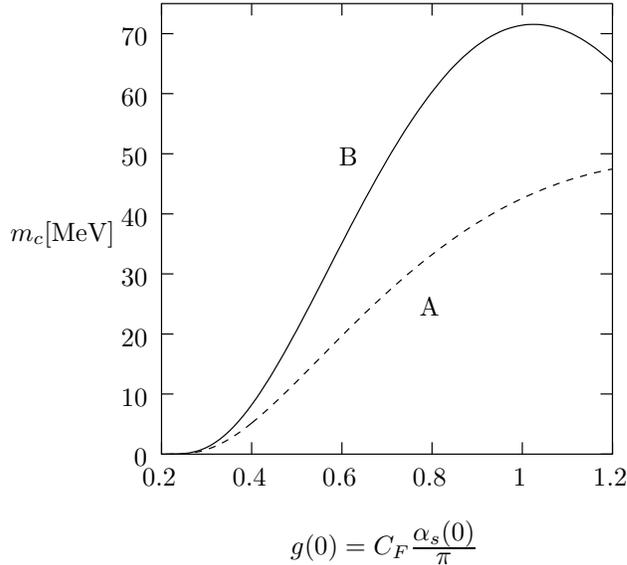}
\end{center}
\caption{Critical mass $m_c$ of the first phase transition 
  as a function of the coupling with behaviour of 
  type A and B, respectively}
\label{fig:mc2g}
\end{figure}

Fig.\ \ref{fig:mc2g} shows the dependence of the first 
critical mass $m_c$ on the coupling strength in models 
A and B. 
In order to obtain this figure we have computed curves 
similar to the one in Fig.\ \ref{fig:m2m} for different 
values of the coupling and determined the turning point 
corresponding to $m_c$. 
Below the critical coupling there are no phase transitions 
and the critical mass acquires values different from zero 
only for supercritical values of the coupling. 
As expected, the critical mass then grows with increasing 
coupling. Realistic values of the coupling are roughly 
of the order $\alpha_s(0) \simeq 0.8$ (model A), 
corresponding to $g(0) \simeq 0.3$. From Fig.\ \ref{fig:mc2g} 
we see that in this region both models give very similar 
values for the critical mass $m_c$ in the range of a few MeV. 
Only for perturbative quark masses below this small 
$m_c$ we expect chiral symmetry breaking. 
This is in perfect agreement with Gribov's physical picture 
according to which exactly the existence of very light quarks 
with masses below of few MeV leads to a dramatic change 
in the vacuum structure. 

At very large values of the coupling there is a maximum 
and the critical mass decreases again. For model B this 
maximum is reached at $g(0) \simeq 1.1$, and for model  
A it is found at $g(0) \simeq 1.5$ (outside the range 
shown in Fig.\ \ref{fig:mc2g}). 
This seemingly strange behaviour has its origin in a 
property of the equation that we notices already in sections 
\ref{asymptbe}. 
There we found that the oscillations in the solutions 
$\chi$ around the value $\pi$ 
disappear at very large coupling $g>1+\sqrt{2/3}$. 
Had we chosen a fixed value for $\alpha_s$ instead 
of a running $\alpha_s$ in Fig.\ \ref{fig:mc2g} we would have 
found in fact that the critical mass vanishes again at 
$g=1+\sqrt{2/3}$. But in our models the running coupling 
is a continuous function. 
Even for arbitrarily large $g(0)$ it has in some momentum range 
values which make oscillations possible, 
and $m_c$ does not vanish even 
at large coupling. Obviously, this is true for 
all possible continuous shapes of the running coupling. 
We therefore expect that the occurrence of chiral symmetry 
breaking due to this mechanism is largely independent of 
the details of the running coupling at small momenta. 

\begin{figure}[htbp]
\begin{center}
\input{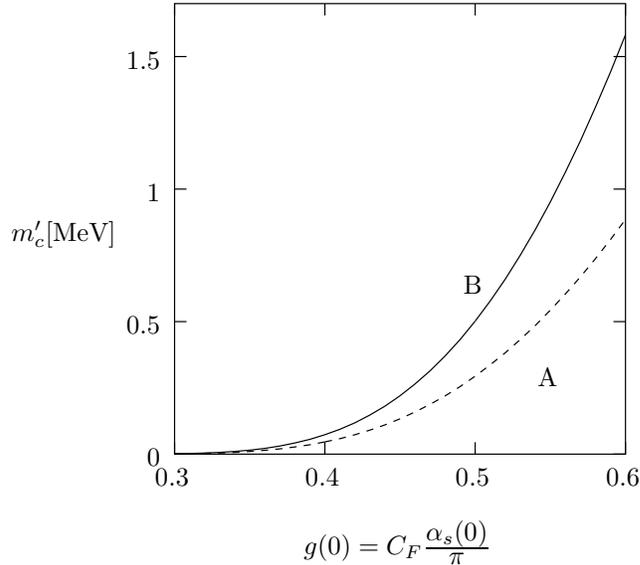}
\end{center}
\caption{Critical mass $m_c'$ of the second phase transition
   as a function of the coupling with behaviour of 
   type A and B, respectively}
\label{fig:mcstrich2g}
\end{figure}

In Fig.\ \ref{fig:mcstrich2g} we show the second 
critical mass $m_c'$ as a function of the coupling. 
As we already mentioned the values of $m_c'$ are almost 
two orders of magnitude smaller than the corresponding 
values of the first critical mass $m_c$. The behaviour 
of $m_c'$ at very large coupling (not shown in the figure) 
is analogous to the behaviour observed for the first 
critical mass, and there is a similar maximum and 
decrease at very large $g(0)$. 

Finally, we would like to comment on the solutions of 
the second and third class found in section \ref{solutionseucreg}. 
Here the interpretation in terms of a renormalized mass 
is more involved than in the solutions of the first class 
discussed above. 
This is because here we find $\chi(0)= -\pi$ (second class) 
or $\chi(0)= \pi$ (third class), respectively. According 
to (\ref{Massenfkt}) this implies that a renormalized 
mass defined as the limit of the mass function at zero 
momentum, see (\ref{defmR}),  would always vanish. 
Nevertheless, it should be possible to define a quantity 
similar to the renormalized mass $m_R$ at some 
intermediate but small momentum scale. 
In the second and third class of solutions 
the behaviour at large momenta is similar to that of 
the first class, especially concerning the oscillations. 
Given a suitable definition of a renormalized mass in 
the above sense one would therefore observe 
similar phase transitions and the breaking of chiral 
symmetry. A full interpretation of the  
renormalized mass, especially for the second and third class 
of solutions, remains to be found. Most probably 
this would require a better understanding of the difficult 
problem of the emergence of a constituent mass of the 
quark. 

\section{Analytic structure of the solutions}
\label{analyticstruct}

So far we have discussed the properties of the solutions 
of Gribov's equation in the region of space--like momenta. 
Now we turn to the problem of determining 
the corresponding analytic structure of the Green 
function in the whole momentum plane. 
The locations and nature of the singularities in the complex 
$q^2$ plane contain crucial information about the solutions, 
especially about their properties regarding causality and 
confinement. It has been argued in \cite{EPJ2} that 
a suitable solution for the retarded Green function of 
a confined quark should be free of singularities in the 
physical region $\imag q^2 \ge 0$. 
The analysis of the first class of solutions (see 
section \ref{solutionseucreg}) in \cite{EPJ1} has shown 
that we should not expect to find such a confining 
solution of the equation without pion corrections. 
We will confirm that result and extend it  
also to the two other classes of solutions. 

As a first step we need to define the coupling 
constant in the whole complex momentum plane. 
This problem, however, is poorly understood even 
at large momenta since the one--loop formula for 
the running coupling does not possess a unique 
analytic continuation. Obviously, the situation is 
even worse at low $|q|$. In order to avoid 
additional parameters and artificial singularities 
we will therefore make 
the simplest possible assumption and choose a 
fixed coupling throughout the complex $q$-plane. 
The value of this coupling will be chosen to be 
supercritical, and in the examples below $\alpha_s=0.8$. 
We have to hope that this choice will be 
a suitable approximation at least for small $|q|$. 
For large $|q|$ our choice seems less appropriate. But 
it will turn out that the singularities of the Green function 
occur at comparatively small values of $|q|$, 
probably indicating that an adiabatic modification of the coupling 
at large $|q|$ would not change the analytic structure. 
Obviously, the choice of a constant and supercritical 
coupling disagrees with the models for $\alpha_s$ 
we have used to find the solutions in the Euclidean region. 
The details of the solutions will in fact change at large 
space--like momenta. But we have checked that the 
classification of the solutions given in section 
\ref{solutionseucreg} remains valid. 

In order to study the analytic structure we 
integrate the equation along suitable contours in the 
complex plane. These should start in the 
Euclidean region where the necessary conditions on 
the initial values are known and 
the possible solutions have been classified (see 
section \ref{solutionseucreg}). 
Some typical contours 
are shown in Fig.\ \ref{fig:Wege}. 
\begin{figure}[htbp]
\begin{center}
\input{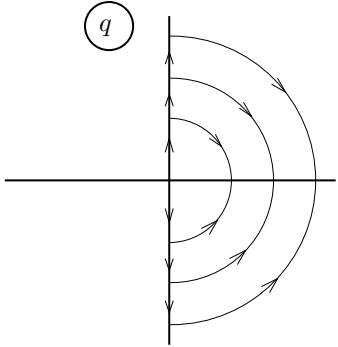}
\end{center}
\caption{Typical contours in the $q$-plane}
\label{fig:Wege}
\end{figure}
In general (especially for the solutions of the second 
and third class) the contours need not start in the 
origin of the $q$-plane. 
We should remind the reader that our definition 
(\ref{defofq}) implies that the right--hand half of the 
$q$-plane already covers the full $q^2$-plane, and 
the Euclidean region is given by the vertical axis in 
Fig.\ \ref{fig:Wege}. 

The occurrence of poles and cuts in the analytic 
continuation will in general lead to a complicated 
Riemann surface. We will in the following only 
discuss singularities arising in the physical region 
of the $q$-plane, 
although it is in principle possible to locate also 
further singularities on unphysical sheets of the 
Riemann surface. 

\begin{figure}[htbp]
\begin{center}
\input{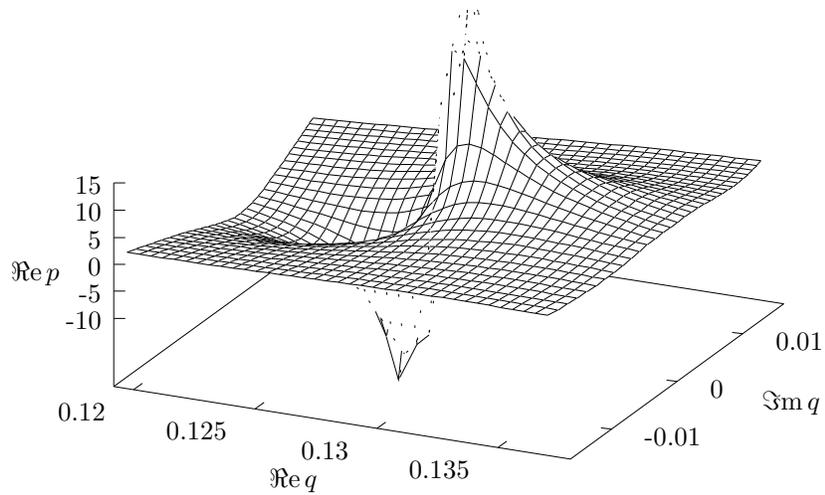}
\end{center}
\caption{Pole of $\real p$ on the real axis, 
  continuation of the solution in Fig.\ \protect\ref{fig:pert}}
\label{fig:phi1pol2}
\end{figure}

We start with the solutions of the first class for which 
$\phi(q=0)=0$ and $\phi \to i \pi$ for large space--like 
$q$, see Fig.\ \ref{fig:pert} for a typical example. 
The different solutions of this class are parametrized 
by the renormalized mass $m_R$. All of these solutions  have a 
pole on the real (time--like) $q$-axis and a cut along this axis 
starting at the pole. There are no other singularities in 
the physical region of the $q$-plane\footnote{There are 
further poles on the unphysical sheets of the cut 
which we will not discuss here.}. This result holds for 
all solutions of this class independent of their renormalized 
mass. In particular, this structure is found even for solutions 
which are located on different branches in 
Fig.\ \ref{fig:m2m}. The singularity and 
the cut on the real axis have been discussed in \cite{Lund,EPJ1} 
but there the absence of further singularities could not be 
proved. To illustrate our result we show in 
Fig.\ \ref{fig:phi1schnitt} the continuation of the solution 
of Fig.\ \ref{fig:pert}, i.\,e.\ in this solution again 
$m_R =100 \,\mbox{MeV}$. 
\begin{figure}[htbp]
\begin{center}
\input{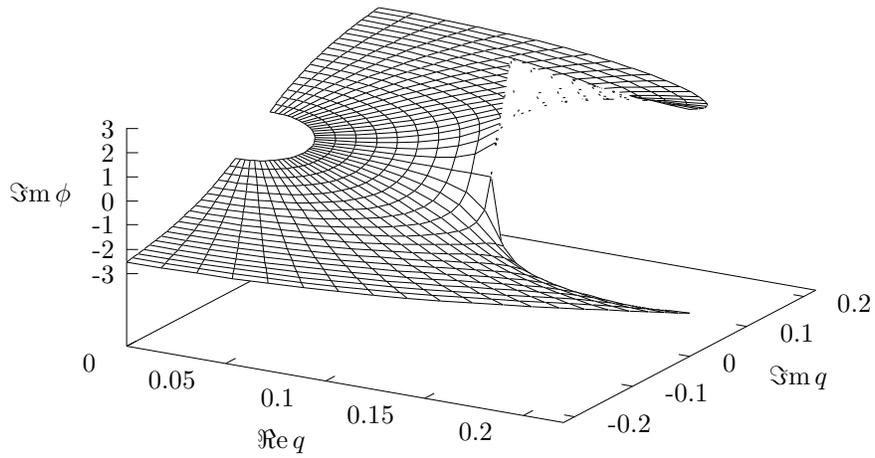}
\end{center}
\caption{Continuation of $\imag \phi$ of the solution in  
  Fig.\ \protect\ref{fig:pert} with a cut on the real $q$-axis }
\label{fig:phi1schnitt}
\end{figure}
In this and in the following figures the momentum $q$ is 
given in units of GeV. The cut is clearly visible. 
Below the cut the function $\phi$ is real--valued on the 
real axis. The starting point of the cut coincides with a 
pole of the Green function at $q=m^*$, 
and in our example $m^*=128\,\mbox{MeV}$. 
The pole is most easily seen in the function $p$, the real part of 
which is plotted in Fig.\ \ref{fig:phi1pol2} in a small region 
around the pole. 
At this point the function $\rho$ (see eq.\ (\ref{Parametrisierung})) 
vanishes such that the Green function develops a pole. 
In general, the pole on the real axis moves to larger values 
of $m^*$ when $m_R$ is increased. 

This analytic structure of the Green function in the first 
class of solutions resembles the analytic structure 
in perturbation theory. The position of the pole can 
be interpreted as the mass $m^*$ of a propagating 
particle. Obviously, these solutions do not lead to 
confined quarks. 

\begin{figure}[htbp]
\begin{center}
\input{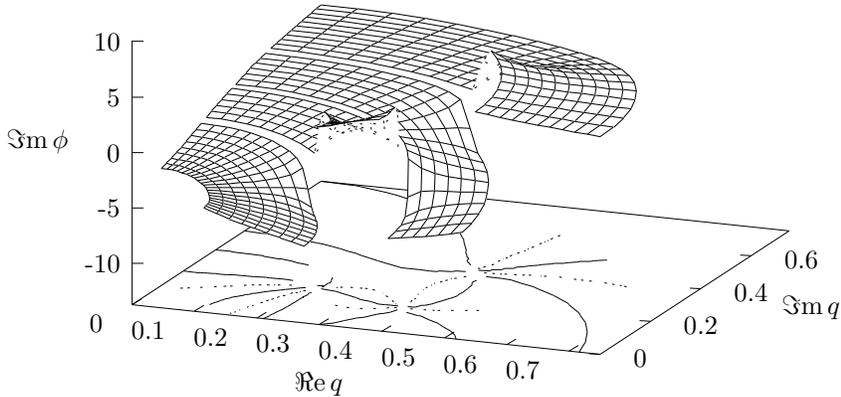}
\end{center}
\caption{$\imag \phi$ in the physical region, 
  continuation of the solution in 
  Fig.\ \protect\ref{fig:adrueber}}
\label{fig:adrueberschnitt}
\end{figure}

Let us now consider the solutions of the second class 
in which $\phi$ runs from $-i\pi$ to $i\pi$ in the 
Euclidean region. These solutions have been obtained 
in section \ref{solutionseucreg} by choosing 
at some point $q_A=i r$ on the imaginary $q$-axis 
the initial values $\phi=0$ and $p=0$. (This 
is the `symmetric' case, for the other case see below.) 
We find that in all solutions of this kind 
the Green function develops a pole exactly on 
the circle with radius $r$, with a cut starting 
at the pole. The exact position of the 
pole depends on the value chosen for $\chi'(q_A)$. 
(In this section we give this parameter of the solutions 
in terms of $\chi$ rather than in terms of $\phi$ 
since $\chi'(q_A)$ is real--valued and in order to 
make it easier to compare with the preceding section. 
We recall that 
the functions are related by $\phi=i \chi$ and 
the derivative is with respect to $\qt=-iq$, see eq.\ 
(\ref{defqprime}).)
For small values of this parameter the pole is 
close to the real $q$-axis, and can in fact be shifted 
arbitrarily close to the real axis by choosing 
sufficiently small $\chi'(q_A)$. For larger values of 
this parameter the pole moves along the circle 
towards the imaginary $q$-axis. In all cases the pole 
is found in the physical region above the real axis. 
In addition, there are two further poles and cuts in the 
physical region. These two poles are located under 
the same polar angle in the $q$-plane. For smaller 
values of $\chi'(q_A)$ they are located closer to the 
real axis, and the one located at smaller $|q|$
moves closer to the origin, whereas the other one moves 
to larger distances from the origin. 
At all three poles the function $\rho$ vanishes. 
As an example we show in Fig.\ \ref{fig:adrueberschnitt} 
the function $\imag \phi$ for the same solution 
which is shown 
in Fig.\ \ref{fig:adrueber} in the Euclidean region, 
i.\,e.\ for the solution obtained from 
$r=0.4\,\mbox{GeV}$ and $\chi'(q_A)=10$. 
The shape of the cuts in the figure is due to the choice 
of the contours of integration. The corresponding poles 
can be found again by looking at the function $p$. 
Their positions can be seen in Fig.\ \ref{fig:adrueber} 
as the points where the contour lines meet. 
There are further singularities in the unphysical region 
which we do not discuss here. 

There are further solutions in the second class which 
we called `asymmetric' in section \ref{solutionseucreg} 
in which the zeros of $\phi$ and $p$ in the Euclidean 
region do not coincide. Also these solutions have a 
singularity in the physical region on the circle on which 
$p=0$ is chosen on the imaginary $q$-axis. As in the `symmetric' 
solution this pole can be shifted very close to the real axis but not 
below it. The `asymmetric' solutions also have further 
singularities but we will not discuss them here. 

\begin{figure}[htbp]
\begin{center}
\input{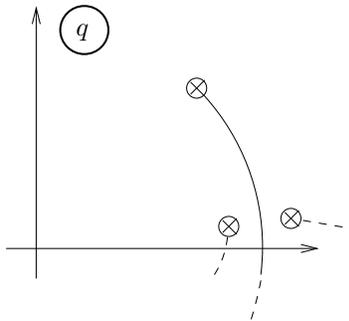}
\end{center}
\caption{Positions of the singularities in the physical region 
  of the $q$-plane for the continuation of the solution in 
  Fig.\ \protect\ref{fig:adrin}}
\label{fig:adrinpole}
\end{figure}

Finally, there is the third class of solutions in which $\phi$ 
stays close to $i \pi$ throughout the Euclidean region. 
Here the situation is similar to the one for the solutions of the 
second class. We find a pole and a cut on the circle on 
which on the imaginary $q$-axis $p=0$. Again this 
is the case for the `symmetric' solutions (in which 
$p=0$ coincides with $\phi=i \pi$ on the imaginary 
axis) as well as for the `asymmetric' solutions. 
As an example we present in Fig.\ \ref{fig:adrinpole} 
the analytic structure of the `symmetric' solution that we have
shown in Fig.\ \ref{fig:adrin}. The crosses and lines 
correspond to the poles and cuts. The solid line indicates 
on which sheets of the cut the other singularities are 
located. 
Denoting again as $q_A$ the point on the imaginary 
axis at which $\phi=0$ and $p=0$, we have for this 
solution $q_A=i \cdot 0.5 \,\mbox{GeV}$ and 
$\chi'(q_A)=4$. For smaller values of $\chi'(q_A)$ 
the pole moves towards the real axis but stays in the 
physical region. There are two further poles and cuts 
as indicated in the figure. These two poles are located 
on a straight line that goes through the origin. 

In summary, we have found in this section that 
the analytic structure of the solutions of the first class 
is similar to the analytic structure of a perturbative 
Green function. For this class we find a pole and a cut on 
the real $q$-axis and no other singularities in the 
physical region. 
In the other two classes there are poles and cuts on 
the physical sheet. Those solutions will in general not permit 
an interpretation in agreement with causality and 
unitarity and should therefore be regarded as 
unphysical solutions of Gribov's equation. 
We did not find any hint for the existence of 
exceptional solutions in which those singularities 
move to the unphysical sheet for special initial 
conditions. 
We can thus conclude that, as anticipated in 
\cite{EPJ1}, there are no solutions of the equation 
without pion corrections which would lead to 
confined quarks. 

\section{Summary and outlook}

The subject of this paper has been Gribov's equation for 
the Green function of light quarks. We have have outlined 
how this equation is derived from the Dyson--Schwinger 
equation, and we have described how the approximations 
made in the derivation are 
motivated by the physical picture of supercritical color charges 
in QCD. In Gribov's scenario the phenomenon of supercritical 
charges can occur in QCD due to the existence of very light 
quarks and is expected to cause chiral symmetry breaking and 
confinement. In the present paper we have concentrated 
on the equation proposed in \cite{Lund,EPJ1} which 
does not yet contain pion corrections. The equation 
describes the Green function in Feynman gauge. It 
collects the most singular contributions 
to the Dyson--Schwinger equation coming from the 
region of small momenta. At the same time it reproduces 
asymptotic freedom at large momenta. 
The derivation of the equation involves the assumption 
that the running coupling remains finite at small momentum 
scales. This is in agreement with recent phenomenological 
results obtained in the dispersive approach to power corrections 
in QCD. These analyses find a typical value of the coupling 
in the infrared which is above the critical value 
$\alpha_c=0.43$ arising in Gribov's scenario. 

Gribov's differential equation for the light quark's Green 
function is a nonlinear equation and so far only asymptotic 
expansions had been used to study it. We have described 
the corresponding results, especially the fixed point 
structure of the equation resulting from this analysis 
and the emergence of the critical value of the strong coupling. 
We have then performed a complete numerical study  of 
the equation. After defining suitable models for the 
behaviour of the strong coupling constant in the infrared 
region we have classified the possible solutions of the 
equation according to their behaviour in the Euclidean 
region of space--like momenta. At supercritical 
coupling the solutions change their behaviour and 
can oscillate while approaching the fixed points. 
The dynamical mass function of the quark has been 
computed in order to show that this change leads to 
the breaking of chiral symmetry at 
supercritical coupling. Chiral symmetry breaking  
takes place independently of the details of the running 
coupling in the infrared as long as it is supercritical in 
some interval of the momentum. 
The breaking of chiral symmetry is connected with 
the occurrence of a series of phase transitions in the 
vacuum of light quarks. If the perturbative mass $m_P$ 
of the quark, i.\,e.\ the mass defined at a large 
momentum scale, is below a critical mass then there 
are several solutions leading to different quark masses $m_R$ 
at low momentum scales, and in general $m_R$ does not 
vanish even in the limit of vanishing perturbative mass $m_P$. 
We have determined the critical mass as a function of the 
coupling. It turns out to depend considerably on the 
mean value of the coupling in the infrared, but the 
dependence is weaker if the coupling  is only slightly 
above the critical value. In agreement with the physical 
picture of supercritical color charges the phase transitions 
occur only for very small values of the perturbative quark mass. 

In \cite{EPJ1,EPJ2} it has been advocated that a confining 
Green function in Gribov's picture should be free of 
singularities in the physical region including 
time--like momenta. We have therefore studied the 
solutions of Gribov's equation in the whole complex 
momentum plane. 
One class of solutions has an analytic structure similar to 
that of a perturbative Green function, whereas the other 
two classes of solutions have singularities on the physical 
sheet. As anticipated in \cite{EPJ1} there are no solutions 
of the equation studied in the present paper exhibiting 
the structure required for confinement. 

The phase transitions connected with the breaking 
of chiral symmetry lead to the generation 
of pions as Goldstone bosons. Their special properties 
in this approach \cite{EPJ1} indicate that they should 
be taken into account as elementary objects, and that 
the equation for the Green function of the light quark 
should be modified accordingly. 
This modification \cite{EPJ2} is expected to 
change the behaviour of the solutions in the Euclidean region 
only very little. In particular, the breaking of chiral 
symmetry will take place in a very similar way. But the 
pion corrections to the equation are expected to have 
significant effects on the analytic structure of the solutions. 
Due to that it appears possible to find a confining Green function. 
The next step should therefore be a numerical study 
of the corresponding differential equation. 
Although the modified equation with pion corrections 
is more complicated it can be analysed using the same 
methods that have been used in the present paper. 
It would at a later stage also be interesting to 
investigate the effects of an additional scalar 
($\sigma$-)meson which has been widely discussed 
in the literature in the context of meson spectra and of 
sigma models.  

We hope that the results of our analysis will be useful 
also for the more conventional approach to the 
Dyson--Schwinger equation for the quark. 
Some general properties of the Green function 
resulting from the Dyson--Schwinger equation 
appear to be rather universal and largely independent of the 
particular approximation scheme that is used. 
The analysis of the Dyson--Schwinger equation 
in quenched supercritical QED \cite{AGW}, 
for example, leads to a picture of phase transitions and 
chiral symmetry breaking that is very similar to the 
one resulting from Gribov's equation. It would 
certainly be useful to identify such universal features 
of the Green function and to achieve a better 
understanding of their origin. 
In this respect it is very important to consider 
different approximation schemes among which Gribov's 
approach is certainly exceptional since it is motivated 
by the very interesting physical picture of 
supercritical charges. 

\section*{Acknowledgements}
This work would not have been possible without numerous 
very instructive and motivating discussions with the 
late Vladimir Gribov. 
It is very sad that he passed away much too early. 
I missed his advice very much during the later stages 
of this work. 

I am most grateful to Yuri Dokshitzer for many very 
helpful discussions and for a careful reading of the 
manuscript, and to Christian Gawron for invaluable advice. 
I would like to thank Bernard Metsch, Julia Nyiri, 
Herbert Petry, Dieter Sch{\"u}tte, Manfred Stingl, 
Bryan Webber, and 
Anthony Williams for helpful discussions. 
I am grateful to the Institut f\"ur Theoretische 
Kernphysik of the University of Bonn where 
part of this work was carried out. Finally, I would like 
to thank the CERN Theory Division for hospitality 
extended to me during the final stages of this work.


\begin{thebibliography}{100}
\bibitem{physica}
V.\,N.\ Gribov, {\it Physica Scripta} {\bf T 15} (1987) 164
\bibitem{Lund}
V.\,N.\ Gribov, 
{\sl Possible Solution of the Problem of Quark Confinement}, 
Lund preprint LU-TP 91-7 (1991)
\bibitem{EPJ1}
V.\,N.\ Gribov, {\it Eur.\ Phys.\ J. }{\bf C 10} (1999) 71 
[hep-ph/9807224]
\bibitem{EPJ2}
V.\,N.\ Gribov, {\it Eur.\ Phys.\ J. }{\bf C 10} (1999) 91
[hep-ph/9902279]
\bibitem{Greiner}
W.\ Greiner, B.\ M\"uller, J.\ Rafelski, 
{\sl Quantum Electrodynamics of Strong Fields}, 
Springer, Berlin, 1985 
\bibitem{Orsay}
V.\,N.\ Gribov, Orsay lectures on confinement (I - III):\\ 
LPT Orsay 92-60, hep-ph/9403218;\\
LPT Orsay 94-20, hep-ph/9404332;\\
LPT Orsay 99-37, hep-ph/9905285
\bibitem{GribovErice} 
V.\,N.\ Gribov, 
in Proceedings of the International School of Subnuclear Physics, 
34th Course, Erice, Italy, 1996, ed.\ A.\ Zichichi, World Scientific, 
p.\ 30
\bibitem{a980}
F.\,E.\ Close, Yu.\,L.\ Dokshitzer, V.\,N.\ Gribov, V.\,A.\ Khoze, 
M.\,G.\ Ryskin, 
{\it Phys.\ Lett.\ }{\bf B 319} (1993) 291 
\bibitem{YuriErice}
Yu.\,L.\ Dokshitzer, 
in Proceedings of the International School of Subnuclear Physics, 
31th Course, Erice, Italy, 1993, ed.\ A.\ Zichichi, World Scientific, 
p.\ 108
\bibitem{YuriVancouver}
Yu.\,L.\ Dokshitzer, 
in Proceedings of 29th International Conference on High 
Energy Physics (ICHEP 98), Vancouver, Canada, 1998, 
p.\ 305 [hep-ph/9812252], 
and references therein
\bibitem{BPY}
Yu.\,L.\ Dokshitzer, G.\ Marchesini, B.\,R.\ Webber, 
{\it Nucl.\ Phys.\ }{\bf B 469} (1996) 93 [hep-ph/9512336] 
\bibitem{power}
Yu.\,L.\ Dokshitzer, B.\,R.\ Webber, 
{\it Phys.\ Lett.\ }{\bf B 352} (1995) 451 [hep-ph/9504219]; 
{\it Phys.\ Lett.\ }{\bf B 404} (1997) 321 [hep-ph/9704298]; 
\\
M.\ Beneke, V.\,M.\ Braun, V.\,I.\ Zakharov, 
{\it Phys.\ Rev.\ Lett.\ }{\bf 73} (1994) 3058;\\
M.\ Beneke, V.\,M.\ Braun, 
{\it Phys.\ Lett.\ }{\bf B 348} (1995) 513 [hep-ph/9411229];\\
P.\ Ball, M.\ Beneke, V.\,M.\ Braun, 
{\it Nucl.\ Phys.\ }{\bf B 452} (1995) 563 [hep-ph/9502300];\\
R.\ Akhoury, V.\,I.\ Zakharov, 
{\it Phys.\ Lett.\ }{\bf B 357} (1995) 646 [hep-ph/9504248]; 
{\it Nucl.\ Phys.\ }{\bf B 465} (1996) 295 [hep-ph/9507253];\\
 G.\,P.\ Korchemsky, G.\ Sterman, 
{\it Nucl.\ Phys.\ }{\bf B 437} (1995) 415 [hep-ph/9411211];
in Proc.\ 30th Recontres de Moriond, Meribel les Allues, France, 
1995, J.\ Tran Thanh Van (ed.) (Editions Frontiers, 1995) 
[hep-ph/9505391] 
\bibitem{Martin}
M.\ Beneke, {\it Phys.\ Rep.\ }{\bf 317} (1999) 1 
[hep-ph/9807443]
\bibitem{DSreview}
C.\,D.\ Roberts, A.\,G.\ Williams, 
{\it Prog.\ Part.\ Nucl.\ Phys.\ }{\bf 33} (1994) 477 
[hep-ph/9403224] 
\bibitem{Bryanalpha}
B.\,R.\ Webber, 
{\it J. High Energy Phys.\ }{\bf 10} (1998) 012 [hep-ph/9805484], 
and references therein
\bibitem{Shirkov}
D.\,V.\ Shirkov and I.\,L.\ Solovtsov, 
{\it Phys.\ Rev.\ Lett.\ }{\bf 79} (1997) 1209 [hep-ph/9704333] 
\bibitem{Buras}
A.\,J.\ Buras, {\it Rev.\ Mod.\ Phys.\ }{\bf 52} (1980) 199
\bibitem{colsys}
U.\ Ascher, J.\ Christiansen , R.\,D.\ Russell,
{\it Math.\ Comp.\ }{\bf  33} (1979) 659 
\bibitem{Bass}
S.\,D.\ Bass, 
{\it Phys.\ Lett.\ }{\bf B 329} (1994) 358 [hep-ph/9404294]
\bibitem{AGW}
F.\,T.\ Hawes, T.\ Sizer, A.\,G.\ Williams, 
{\it Phys.\ Rev.\ }{\bf D 55} (1997) 3866 [hep-ph/9608292] 
\end{thebibliography}
\end{document}